\documentclass[aps,floats,floatfix,amssymb,amsmath,prb,twocolumn,superscriptaddress]{revtex4-2}

\usepackage{calc}
\usepackage{psfrag}
\usepackage{graphicx}
\usepackage{soul,color}
\usepackage[dvipsnames]{xcolor}
\usepackage[unicode=true,pdfusetitle,
 bookmarks=true,bookmarksnumbered=false,bookmarksopen=false,
 breaklinks=false,pdfborder={0 0 0},backref=false,colorlinks=true]
{hyperref}
\usepackage{geometry}
\usepackage{float}

 \setstcolor{blue}



\begin{document}

\title{Thermal conductivity and heat diffusion in the two-dimensional Hubbard model}

\author{Martin Ulaga}
\affiliation{Jo\v{z}ef Stefan Institute, Jamova 39, 1000 Ljubljana, Slovenia}
\author{Jernej Mravlje}
\affiliation{Jo\v{z}ef Stefan Institute, Jamova 39, 1000 Ljubljana, Slovenia}
\author{Peter Prelov\v sek}
\affiliation{Jo\v{z}ef Stefan Institute, Jamova 39, 1000 Ljubljana, Slovenia}
\author{Jure Kokalj}
\affiliation{University of Ljubljana, Faculty of Civil and Geodetic
  Engineering, Jamova 2, 1000 Ljubljana, Slovenia} 
\affiliation{Jo\v{z}ef Stefan Institute, Jamova 39, 1000 Ljubljana, Slovenia}

\begin{abstract}

We study the electronic thermal conductivity $\kappa_\textrm{el}$ and the
  thermal diffusion constant $D_\textrm{Q,el}$ in the square lattice Hubbard
  model using the finite-temperature Lanczos method. We exploit the
  Nernst-Einstein relation for thermal transport and interpret the
  strong non-monotonous temperature dependence of $\kappa_\textrm{el}$ in
  terms of that of $D_\textrm{Q,el}$ and the electronic specific heat
  $c_\textrm{el}$. We present also the results for the Heisenberg model on a
  square lattice and ladder geometries. 
  We study the effects of doping and consider
  the doped case also with the dynamical mean-field theory. We show 
  that $\kappa_\textrm{el}$ is below the corresponding  Mott-Ioffe-Regel value in 
  almost all calculated regimes, while the mean free path is typically above or close to lattice spacing.
  We discuss the opposite effect of quasi-particle renormalization on charge and heat diffusion constants. We calculate the
  Lorenz ratio and show that it differs from the Sommerfeld value. We discuss our results in relation to experiments on cuprates.  Additionally, we calculate the thermal conductivity of overdoped cuprates within the anisotropic marginal Fermi liquid phenomenological approach.
\end{abstract}
\maketitle

\section{Introduction}

Thermal conductivity is a powerful probe of correlated electrons which, e.g., allowed detection of the breakdown of the Fermi-liquid theory in cuprate superconductor \cite{hill01}, observing highly mobile excitations in organic spin liquid \cite{yamashita10} and determining the absence of quasiparticles in electronic fluid of vanadium dioxide \cite{lee17}.
Despite this, thermal conductivity receives less attention than the charge conductivity, 
which was recently measured also in optical lattices~\cite{brown19}, and was explored theoretically with precise numerical simulations both in the high-temperature bad metal~\cite{kokalj17,perepelitsky16, brown19,vucicevic19,vranic20} and lower temperature strange metal regime~\cite{huang19}. 

Both cold atom measurements~\cite{brown19} and theoretical discussions of transport properties employ the Nernst-Einstein relation that expresses the conductivity   $\sigma_\textrm{c}=\chi_\textrm{c} D_\textrm{c}$ in terms of the charge susceptibility $\chi_\textrm{c}$ and the charge diffusion constant $D_\textrm{c}$. At high temperatures the temperature dependence of $\sigma_\textrm{c}$ is dominated by $\chi_\textrm{c}$ \cite{kokalj17,perepelitsky16, brown19} and one can understand the appearance of bad-metallicity (conductivity below the Mott-Ioffe-Regel value~\cite{gunnarsson03}) in terms of decreasing $\chi_\textrm{c}$ and saturated, temperature independent $D_\textrm{c}$. 
Similarly, one can express the spin
conductivity $\sigma_\textrm{s}=\chi_\textrm{s} D_\textrm{s}$ (with $\chi_\textrm{s}$ being the uniform spin
susceptibility and $D_\textrm{s}$ being the spin diffusion constant). This was used in
a study of spin transport in cold atoms \cite{nichols19}.
The spin diffusion constant has a non-monotonic $T$-dependence and reaches values
below the lower limit of charge diffusion. This occurs because the 
velocity is reduced from a value given by hopping $t$ to a lower one given
by the (lower energy) Heisenberg exchange $J$ \cite{nichols19,ulaga21}. 

One can also discuss the thermal conductivity along the lines of the corresponding Nernst-Einstein relation
$\kappa=c D_\textrm{Q}$ (with $c$ being the specific heat and $D_\textrm{Q}$ the heat
diffusion constant). In contrast to the case of charge conductivity,  
$\kappa$,
$c$ and $D_\textrm{Q}$ can all be independently measured~\cite{zhang17,zhang19,martelli18}. 
One could thus hope  
for better characterization of the electronic
transport, but $\kappa=\kappa_\textrm{ph}+\kappa_\textrm{el}$  has both electronic and
phononic contributions and separating them is not straightforward. 
One typically resorts to estimating 
electronic contribution $\kappa_\textrm{el}$  via the
Wiedemann-Franz law, which, however is often violated~\cite{lee17,zhang00,kim09,mahajan13,lavasani2019}. 
The difficulty
to unambiguously identify the two contributions can be illustrated in the case of 
the normal state in cuprates, where one can find quite
diverse claims:
(i) $\kappa_\textrm{el}$ represents about a half of the total
$\kappa$ \cite{yu92,allen94} or (ii) a very
small portion of total $\kappa$ \cite{yan04, zhang19} (iii)
which contrasts with a surprisingly large magnonic contribution found in Ref.~\onlinecite{hess03},
and (iv) total $\kappa$ showing the same in-plane
anisotropy as $\sigma_\textrm{c}$, suggesting it has an electronic origin
\cite{zhang17}. Recent studies discuss the phononic part $\kappa_\textrm{ph}$
in terms of a Planckian relaxation rate \cite{zhang19, mousatov20},
but they could still be affected by the uncertainties in the  subtraction of the electronic part.

Is the behavior of $\kappa_\textrm{el}$ better characterized at least within theory? 
Thermal transport was broadly studied in one-dimensional systems in part due to much larger values of $\kappa$ originating in long mean free paths and proximity to integrability~\cite{zotos98,zotos05, hess07,karrasch17}. Results for dimension $d>1$ are however scarce.  
The Hubbard model in 2 dimensions was very recently studied with a determinant quantum Monte Carlo investigation of the Mott insulator \cite{wang21} and with a weak coupling approach \cite{kiely21}, but no other results exist. We are unaware of any calculation of thermal conductivity even for the more basic 2d Heisenberg model. It is important to have robust numerical results for $\kappa_\textrm{el}$ not only to address the fundamental questions, e.g., asymptotic behavior of the diffusion constant and relaxation rates, but also to help interpreting the experiments. 

In this work we study the thermal conductivity and the heat diffusion constant
in the square lattice Hubbard model with the finite temperature Lanczos method (FTLM). 
We study also the Heisenberg model both on square lattice and ladder geometries.
Our FTLM calculations are limited to $T\gtrsim J/2$ and thus map out the high temperature regime of the phase diagram and are directly relevant for the cold atom experiments \cite{brown19,nichols19,schneider12, gardadosanchez19}. There the
density, spin density and energy density relaxations are affected also
by a thermal conduction either directly or via mixed, e.g.,
thermoelectric effects \cite{mravlje22}.
For materials, the experimental temperatures are usually lower. 
To discuss this regime we thus resort to a phenomenological spin-wave model, qualitative aspects of the dynamical mean field theory (DMFT) results, and the phenomenological anisotropic marginal Fermi liquid model (AMFL). The latter captures various aspects of overdoped cuprates 
 \cite{kokalj11,kokalj12}. We discuss our results for the Mott-insulating and doped cases in relation to experiments on cuprates. 

The paper is structured as follows. In Sec.~\ref{sec_model} we present models and methods.
We show the results for the Mott insulating state with strongly nonmonotonic behavior of  $\kappa_\textrm{el}$ in Sec.~\ref{sec_mott}, where we also discuss the difference between the Hubbard and Heisenberg model results. 
The effect of doping is presented in Sec.~\ref{sec_doped}  and the violation of the Wiedemann-Franz law is discussed in Sec.~\ref{sec_wf}. 
In Sec.~\ref{sec_exper}, our results are discussed in relation to experiments on cuprates for undoped and doped regime. We summarize  our findings in Sec.~\ref{sec_concl}. Appendix~\ref{app-ftlm} contains technical details of FTLM calculations. We discuss cluster shape dependence of the results in Appendix~\ref{app-cluster}, frequency dependence of conductivity in Appendix~\ref{app-freq}, vertex corrections in Appendix~\ref{app-vertex}, the quasiparticle regime in Appendix~\ref{app-quasi} and the AMFL phenomenology for overdoped cuprates in Appendix~\ref{app-amfl}.

\section{Model and method}
\label{sec_model}
We consider the Hubbard  model on a square lattice,
\begin{equation}
 H = -t\sum_{\langle i,j\rangle,s} c^\dagger_{i,s} c_{j,s}+U\sum_{i}
 n_{i,\uparrow}n_{i,\downarrow},
\label{eq_ham}
\end{equation}
where $c^\dagger_{i,s}/c_{i,s}$ create/annihilate an electron with spin
$s$ (either $\uparrow$ or $\downarrow$) at the lattice site $i$. The hopping amplitude between the nearest neighbors is $t$. We further set $\hbar=k_\textrm{B}=e=1$. 
We denote the lattice parameter with $a$.

We investigate the model with FTLM
\cite{jaklic00,prelovsek13,kokalj13} on a cluster with size $N=4\times 4$. 
To reduce the finite-size effects that appear at low $T$,
we employ averaging over twisted boundary conditions and use
the grand canonical ensemble. We do not show 
results in the low $T$ regime where our estimated 
uncertainty due to finite size
effects and spectra broadening exceeds 20\%.
We also perform FTLM calculations on the square lattice Heisenberg model with up to 32 sites and additionally on 2 leg and 3 leg ladders. 
For the Hubbard model away from half filling, we additionally compare our results with single-site DMFT calculations, obtained with NRG-Ljubljana~\cite{zitko2009energy,zitko_rok_2021_4841076} as the impurity solver. 

The electronic thermal conductivity is calculated as 
\begin{equation}
\kappa_\textrm{el}=\frac{L_{22}}{T} -\frac{L_{21}^2}{TL_{11}},
\label{eq_kappa}
\end{equation}
where $L_{ij}$ represent corresponding conductivities with
$L_{11}=\sigma_\textrm{n,n}=\sigma_\textrm{c}$, $L_{12}=L_{21}=\sigma_\textrm{Q,n}$ and
$L_{22}=\sigma_\textrm{Q,Q}$.
Within FTLM \cite{jaklic00}, these are 
obtained as the frequency $\omega=0$
value of the dynamical conductivities related to the current-current
correlation functions $\sigma_\textrm{A,B}(\omega)=\textrm{Im}
\chi_\textrm{A,B}(\omega)/\omega$
while in DMFT they are calculated with the bubble approximation.
The heat current $j_\textrm{Q}$ is given by the
energy $j_\textrm{E}$ and particle $j_\textrm{n}$ currents, $j_\textrm{Q}=j_\textrm{E}-\mu j_\textrm{n}$. Within the Heisenberg model $\kappa_\textrm{el}$ is  calculated as $\kappa_\textrm{el}=\sigma_\textrm{E,E}/T$.
More details on  the calculations are given in Appendix~\ref{app-ftlm}.

\section{Mott insulator}
\label{sec_mott}

\begin{figure}[ht!]
 \includegraphics[ width=0.98\columnwidth]{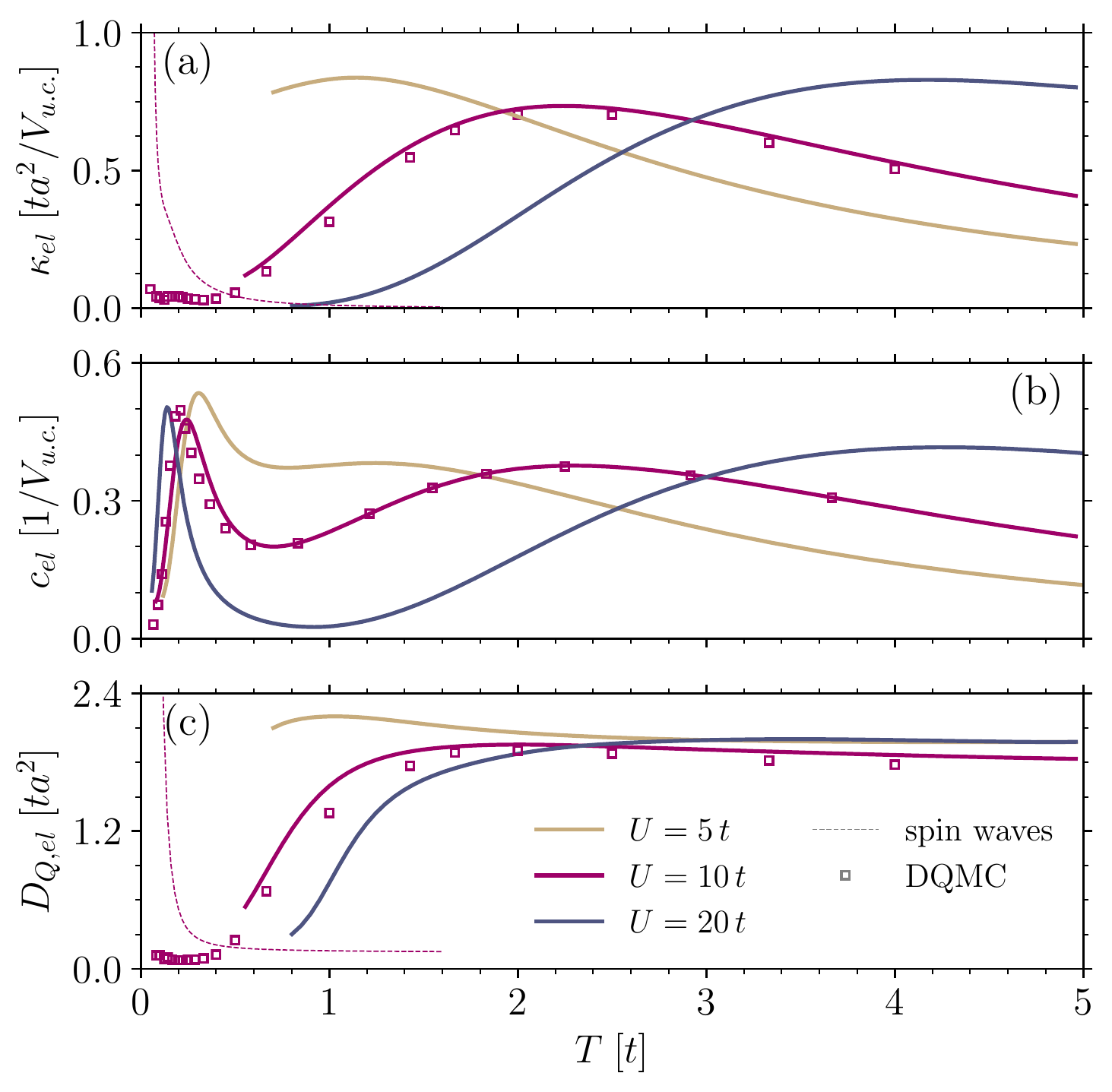} 
 \caption{
The electronic thermal conductivity $\kappa_\textrm{el}$, specific heat $c_\textrm{el}$, and 
thermal diffusion constant $D_\textrm{Q,el}$ for a half-filled (doping $p=0$) Hubbard model with $U=5t$, $10t$ and $20t$. The full thick lines are FTLM results and the squares are DQMC data taken from  Ref. \onlinecite{wang21}.
The thin dashed lines in a) and c) denoted ``spin waves'' are the phenomenological approximation for the spin (Heisenberg) part and are obtained as explained in the main text, fixing $J=4t^2/U=0.4t$.
}
 \label{fig1}
\end{figure}

We show $\kappa_\textrm{el}$ for $U=5t$, $10t$ and $20t$ in the
half-filled ($n=1$, doping $p=1-n=0$) Mott-insulating case in Fig. \ref{fig1}a. The most
prominent feature is the non-monotonic $T$ dependence with a large
maximum at high $T$, e.g., at $T\sim 2t$ for $U=10t$.  This maximum can be understood  via the Nernst-Einstein relation 
$\kappa_\textrm{el}= c_\textrm{el} D_\textrm{Q,el}$ in terms of a maximum in the electronic specific heat $c_\textrm{el}$
\footnote{We use the specific heat at fixed density or doping, which is in contrast with
the specific heat at fixed chemical potential calculated in Refs.~\onlinecite{jaklic00,bonca03,kokalj13}.}. This maximum is
shown in Fig. \ref{fig1}b and 
is the high-$T$ maximum in $c_\textrm{el}$
(opposed to low-$T$ maximum in $c_\textrm{el}$ at $T<t$) and originates in
the increase of the entropy from the spin (Heisenberg) value $\ln(2)$ towards
a full charge activated value $\ln(4)$ via the thermal activation of
mobile doublons and holons \cite{prelovsek15} across the charge
gap $\Delta_\textrm{c}$ \cite{bonca03, kokalj13}. It moves to higher $T$ with
increasing $U$ (see Figs.~\ref{fig1}a,b). The maximum in $\kappa_\textrm{el}$ at high
$T$ is therefore a consequence of the new heat conduction channel via
particles, doublons and holons.

At lower $T$, e.g., $T<2t$ for $U=10t$, $\kappa_\textrm{el}$ 
decreases faster with decreasing $T$ than $c_\textrm{el}$ (see
Figs. \ref{fig1}a,b). This is due to strong decrease of the electronic heat diffusion constant $D_\textrm{Q,el}$ (see Fig.~\ref{fig1}c) and indicates a crossover from 
particle dominated to spin (wave) dominated heat transport at lower $T$ and the accompanying
strong decrease of the average velocity $v$ determining the diffusion
constant $D=v l/2$ \cite{ulaga21}. The velocity $v$ decreases from the order of $v\sim ta$ to the order of $v
\sim Ja$. 
Here $J=4t^2/U$ is the exchange coupling \cite{eskes94} and
$l$ is the mean free path. $D_\textrm{Q,el}$ is calculated 
via the Nernst-Einstein relation $D_\textrm{Q,el}=\kappa_\textrm{el}/c_\textrm{el}$.

\begin{figure}[ht!]
  \includegraphics[ width=0.98\columnwidth]{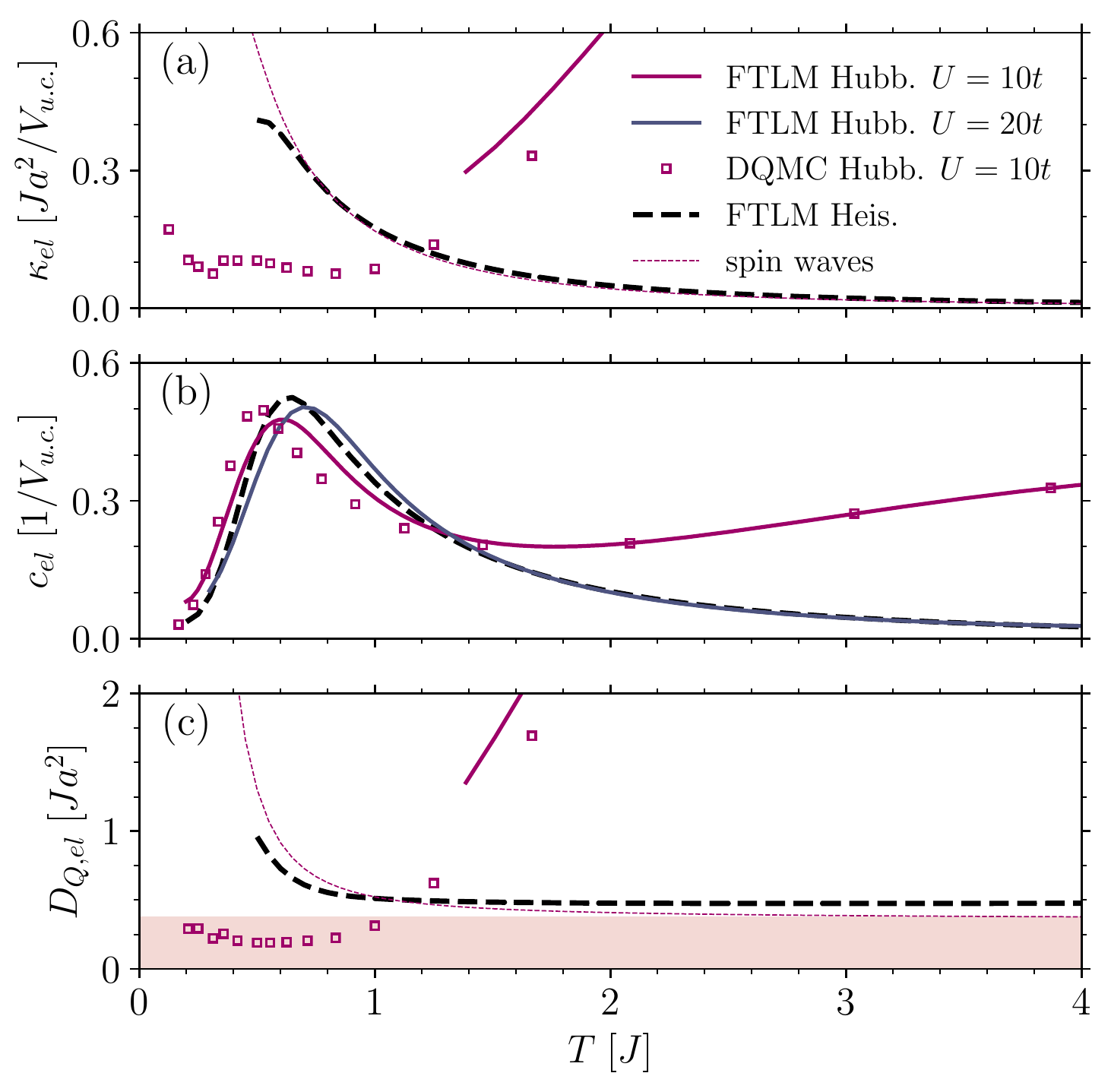} 
 \caption{
$\kappa_\textrm{el}$, $c_\textrm{el}$ and $D_\textrm{Q,el}$ calculated with FTLM and DQMC  
compared to Heisenberg model FTLM ($N=32$) results. The results are shown in units of exchange coupling $J=4t^2/U$. 
Results are compared also to the ``spin waves'' approximation (thin dashed line)
explained in the main text. The shaded area indicates  
the region of spin MIR limit violation. 
DQMC data are taken from  Ref.~\onlinecite{wang21}.
}
 \label{fig_heis}
\end{figure}

At even lower $T\sim J$, $c_\textrm{el}$ shows a peak due to spin excitations
\cite{jaklic00, bonca03,kokalj13}. For large $U$ this peak can be well described with the Heisenberg model as is shown in Fig.~\ref{fig_heis}b. 
Our FTLM Heisenberg results on 32 sites for $c_\textrm{el}$ agree well with the results from 
Refs.~\onlinecite{hofmann03,schnack18} and also reasonably with those from Refs.~\onlinecite{sengupta03,makivic91}, where a peak occurs at a slightly lower $T$.
Whether this peak in $c_\textrm{el}$ manifests as a peak in $\kappa_\textrm{el}$ depends on the strength of the $T$ dependence of $D_\textrm{Q,el}$. If $D_\textrm{Q,el}$ increased strongly with decreasing $T$, the peak in $c_\textrm{el}$ would appear only as a shoulder in $\kappa_\textrm{el}$.

To explore lower $T\lesssim J$ behavior, we calculate $\kappa_\mathrm{el}$ in the Heisenberg model using FTLM. The results are shown in Fig.~\ref{fig_heis} next to the Hubbard model results for $U=10t$. 
The Heisenberg model $\kappa_\textrm{el}$ monotonically increases with
decreasing $T$ (Fig.~\ref{fig_heis}a). This increase becomes less steep
at lowest $T$ that reach below the ones corresponding to the low-$T$
peak in $c_\mathrm{el}$. The diffusion constant is shown on Fig.~\ref{fig_heis}c. It is essentially temperature-independent above $T\sim J $, but increases below it. We studied additional cluster sizes and shapes with
FTLM and report the results in Appendix~\ref{app-cluster}.  Finite
clusters indicate a peak in $\kappa_\textrm{el}$ corresponding to the
peak in $c_\textrm{el}$, but the finite size effects are significant there: with increasing cluster size
$\kappa_\textrm{el}$ is still increasing.

The Heisenberg model results reach values significantly above the DQMC
Hubbard model results for $T<J$. We do not understand this
discrepancy. One could attribute this to the difficulties
associated with analytical continuation in DQMC but note that the
agreement between DQMC and FTLM results for the Hubbard model at higher $T$ is good
(see also Appendix~\ref{app-freq}). The other option could be the higher order corrections in $t/U$ expansion of the Hubbard model.
 
We compare our numerical results also with a phenomenological model. For this, we take $c_\textrm{el}$ from FTLM Heisenberg
model results and approximate 
the mean free path by the spin-spin correlation length $\xi$ from
renormalization group calculations~\cite{chakravarty89,kim98}
\begin{equation}
l=\sqrt{[C_\xi a\exp(2\pi \rho_\textrm{s}/T)/(1+1/(2\pi \rho_\textrm{s} /T))]^2+a^2}.
\end{equation}
$\xi$ is modified to approach $l(T\to \infty) \to a$ and obtained 
with $C_\xi=0.5$ and $\rho_\textrm{s}=0.15J$ and $T$ in units of $J$. 
We approximate the velocity $v$ with the kinetic magnon approximation \cite{notemagnon}, which gives $v\sim 0.72Ja$ at highest $T$ and interpolates to $v\sim 1.4Ja$ at low $T$~\cite{kim98, igarashi05,ulaga21}.
From this, we obtain $D_\textrm{Q,el}=v l/2$ and $\kappa_\textrm{el}=c_\textrm{el}D_\textrm{Q,el}$ and show
them in Fig.~\ref{fig1} and \ref{fig_heis} with thin dashed lines denoted ``spin
waves". The obtained diffusion constant (taking $l=a$) agrees with the known limiting value of the spin diffusion constant $D_s \sim 0.4Ja^2$ in the Heisenberg model at high $T$ \cite{bonca95}. The agreement between the DQMC result and this spin wave estimate in Fig.~\ref{fig_heis}a is poor.

On the other hand, the $T$ dependence of the spin wave estimate and Heisenberg model results agree qualitatively.
$D_\textrm{Q,el}$ in the Heisenberg model at high $T$ is close to the spin-wave estimate (see Fig.~\ref{fig_heis}c). 
At lower $T< J$ one expects $D_\textrm{Q,el}$ to increase (and diverge with $T\to 0$) due to increased (diverging) $l$. It is also expected that the Heisenberg $D_\textrm{Q,el}$ is smaller than the spin wave estimate as observed in Fig.~\ref{fig_heis}c. Namely, one expects $l \lesssim \xi$ since spin waves are expected to scatter at the antiferromagnetic domain walls separated effectively by $\xi$.

\section{Doped Mott insulator}
\label{sec_doped}

\begin{figure}[ht!]
  \includegraphics[width=0.99\columnwidth]{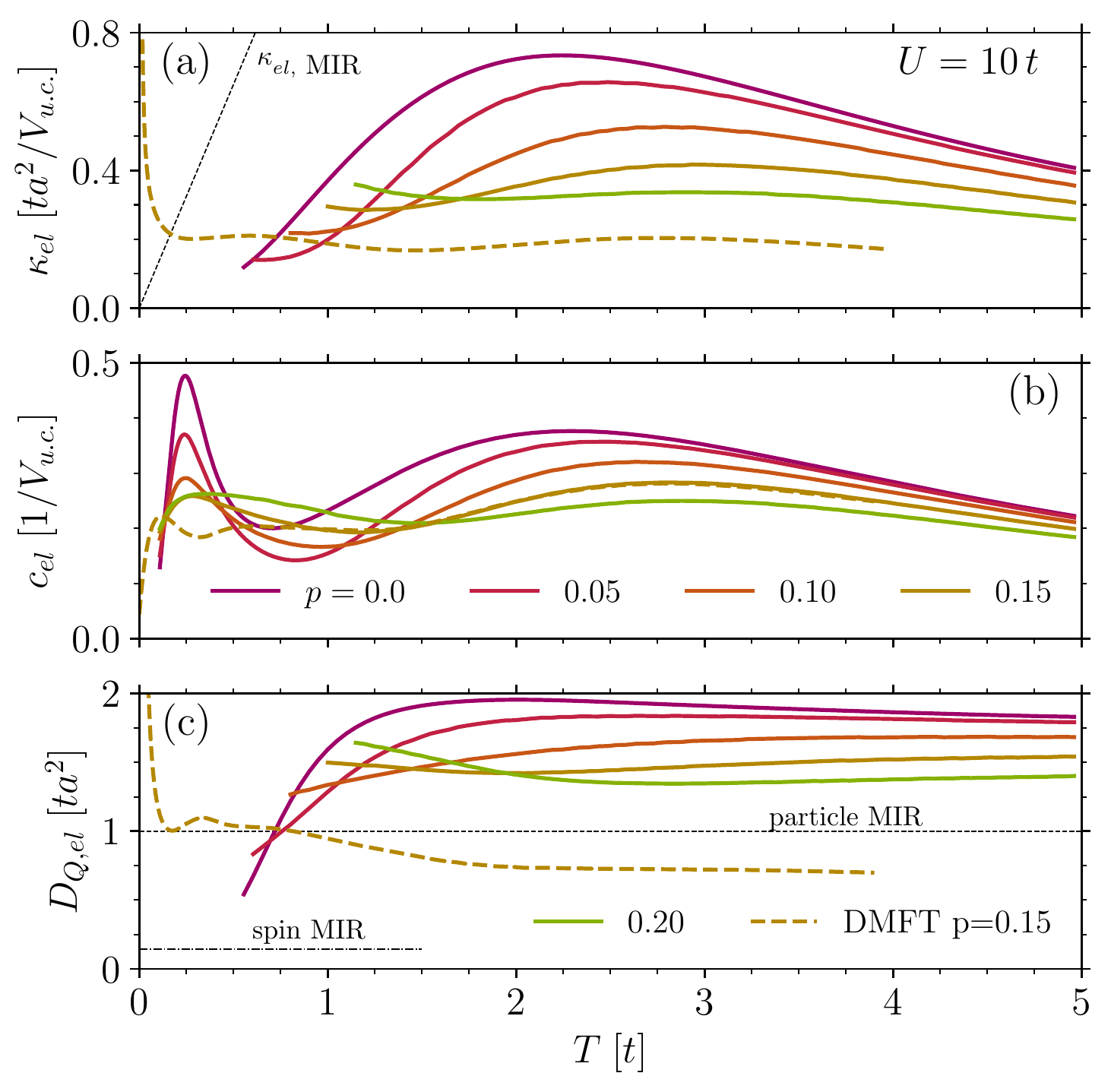}
 \caption{ 
 The temperature dependence of $\kappa_\textrm{el}$ (a), $c_\textrm{el}$ (b), and
   $D_\textrm{Q,el}$ (c) for several hole dopings $p$ and for $U=10t$ as obtained with FTLM.
   DMFT results for $p=0.15$ are also shown. Characteristic MIR limits are indicated in (a) and (c) (see text).
}
 \label{fig2}
\end{figure}

We now consider the effect of doping. We show  $\kappa_\textrm{el}$ for several hole dopings $p$
and for $U=10t$ in Fig.~\ref{fig2}a. With increasing $p$ the high-$T$ peak at $T\sim 2.5t$ becomes
less pronounced, due to suppressed  $c_\textrm{el}$
(Fig.~\ref{fig2}b) and lower release of entropy via thermal
activation of holons and doublons. 
In comparison to the Mott insulating case, $\kappa_\textrm{el}$ at 
$T\lesssim t$ is increased due to charge conduction.
The increase at low-$T$  for larger
dopings indicate the onset of coherence.

\subsection{Mott-Ioffe-Regel limit for $\kappa_\textrm{el}$}

Like for the case of charge conductivity $\sigma_c$, we can introduce the MIR value that indicates the minimal conduction within the Boltzmann estimate by setting $l \sim a$.  In 2d it is given by~\cite{notemir}
\begin{equation}
\kappa_\textrm{el,MIR}= \frac{\pi^2 T}{3} \frac{\sqrt{n}a^2}{\sqrt{2\pi} V_\textrm{u.c.}}.
\end{equation}

With our units and filling $n\sim 1$ one
obtains $\kappa_\textrm{el,MIR} \sim 1.3\,T a^2/V_\textrm{u.c.}$. This value is indicated in
Fig.~\ref{fig2}a and our $\kappa_\textrm{el}$ is well below it. 
It has been shown, that violations of the MIR limit
for the charge conductivity \cite{kokalj17, brown19, perepelitsky16}
(and spin conductivity \cite{ulaga21}) originate in strongly
suppressed static charge susceptibility (or spin susceptibility),
while the diffusion constant and mean free path still correspond to $l\gtrsim a$. 
Is this the case also for $\kappa_\textrm{el}$?

In Fig.~\ref{fig2}c we compare the calculated $D_\textrm{Q,el}$ with the
corresponding MIR value $D_\textrm{Q,el,MIR}=va/2$. The expected velocity in the
doped case is the quasiparticle velocity, which we approximate with
$v\sim 2ta$. This leads to $D_\textrm{Q,el,MIR}\sim
ta^2$. For almost all parameter regimes we observe
$D_\textrm{Q,el}>D_\textrm{Q,el,MIR}$ and $l\gtrsim a$.
At lowest $T$ and half-filling ($p=0$) one expects lower limiting values as the heat conductance is dominated by spins with lower velocity $v\sim Ja$. 
If one uses the spin-wave
velocity $v\sim 0.72Ja$~\cite{notemagnon} one obtains the
lower bound $D_\textrm{Q,el,MIR,spin}\sim 0.36Ja^2$, namely 
$D_\textrm{Q,el,MIR,spin}\sim 0.14ta^2$ for $U=10t$. This value is indicated in
Fig.~\ref{fig2}c and $D_\textrm{Q,el}$ for $p=0$ is above it and only
approaches it with lowering $T$. This is also shown in  Fig.~\ref{fig_heis}c together with the Heisenberg model results, which saturates closely to the MIR value (indicated by shading).

On the other hand, the DQMC results at the lowest $T$ are below the bound and the FTLM results for low doping ($p=0.05$) at the lowest $T$ cross the $D_\textrm{Q,el,MIR}\sim ta^2$ (see Fig.~\ref{fig2}c). Reconciliation of this in terms of a possible deconstruction  $\kappa_\textrm{el}=\kappa_\textrm{spin}+\kappa_\textrm{particle}$ or with effectively decreased velocity from the order of $v\sim ta$ to $v\sim Ja$ remains a subject for future work.

Let us also note that as $T\to 0$ the mean free path is expected to 
diverge, leading to diverging $D_\textrm{Q,el}(T\to 0)\to \infty$ within the Hubbard model. 
Therefore, $D_\textrm{Q,el}$ has a non-monotonic
$T$ dependence. 

To qualitatively explore the behavior at lower $T$, we performed also DMFT calculations for
$p=0.15$ and show the results in Fig.~\ref{fig2}. 
The DMFT results display a 
divergence of $\kappa_\textrm{el}$ and $D_\textrm{Q,el}$ as $T\to 0 $. This rapid growth occurs below $T\sim 0.1t$ due to low coherence
temperature. Remarkably the upturn appears at $T$ where
$\kappa_\textrm{el}\sim \kappa_\textrm{el,MIR}$ (see Fig.~\ref{fig2}a). Similar behavior is observed for charge transport \cite{deng13}.

\subsection{Comparison of heat and charge diffusion}

\begin{figure}[ht!]
 \begin{center}
   \includegraphics[ width=0.99\columnwidth]{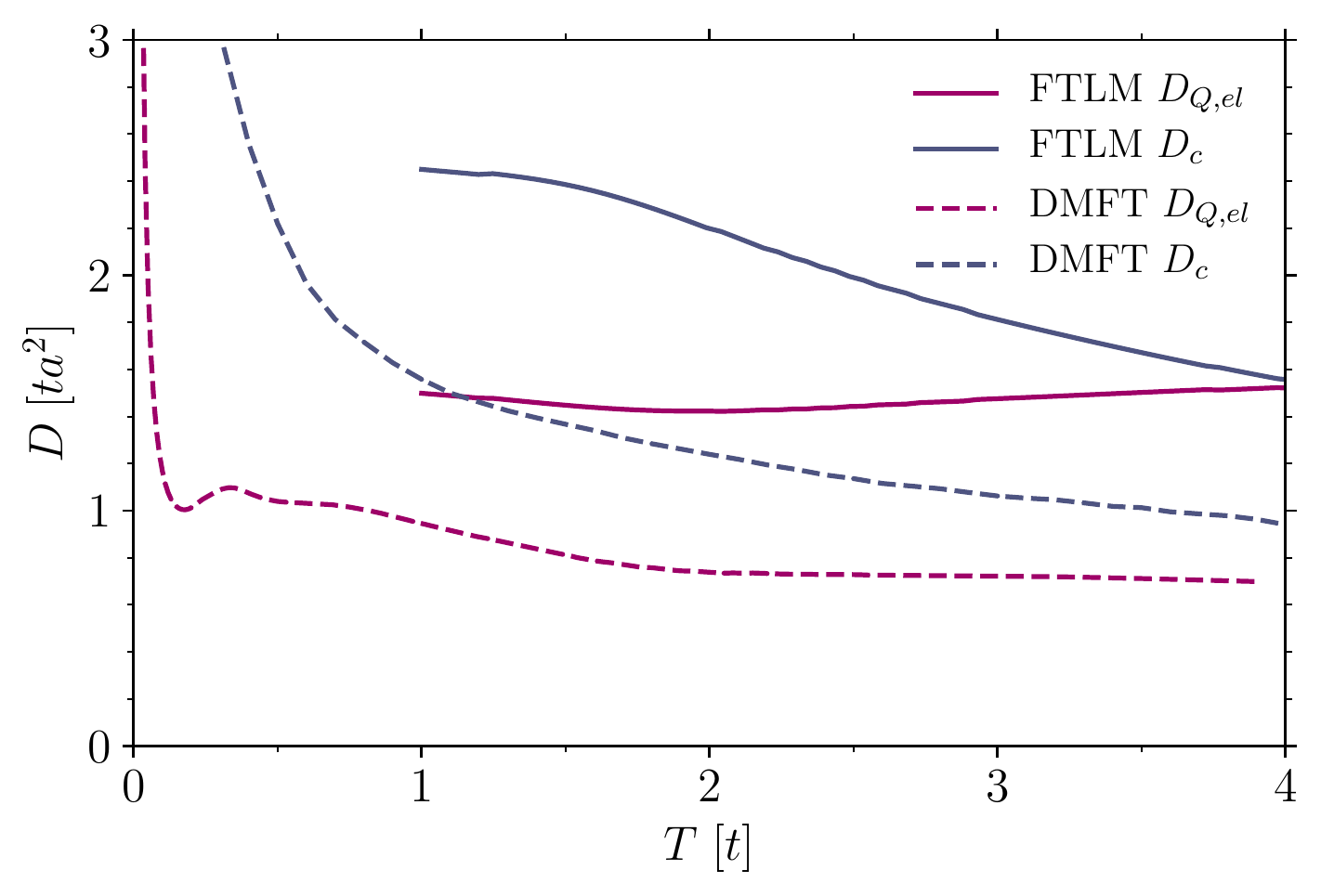}
\end{center}
 \caption{
Comparison of heat and charge diffusion constants $D_\textrm{Q,el}$ and $D_\textrm{c}$ as obtained by FTLM and DMFT. $U=10t$, $p=0.15$.}
 \label{figdqdc}
\end{figure}

For the doped case, it is interesting to compare heat $D_\textrm{Q,el}$ and charge $D_\textrm{c}$ diffusion constants. This is shown in Fig.~\ref{figdqdc} and one can see that $D_\textrm{Q,el}$ and $D_\textrm{c}$ behave differently. The thermal diffusion constant depends on temperature more weakly. One can also see that $D_\textrm{Q,el}$ is smaller than $D_\textrm{c}$ for $T\lesssim 4t$. The heat transport seems less coherent at lower $T$.
Similar trends as indicated by FTLM results continue to higher $T$, where at very high $T$  $D_\textrm{Q,el}\sim 1.6ta^2$ and is larger than $D_\textrm{c} \sim ta^2$. This difference is currently not understood. 

We also show the DMFT result and find that $D_\textrm{Q,el}$ and $D_\textrm{c}$ differ even at low $T$. The difference between FTLM and DMFT result can be attributed to vertex corrections. We discuss this in more detail in 
Appendix~\ref{app-vertex}, where we also show the frequency dependent $\kappa_\textrm{el}(\omega)$ which has a peak at $\omega \sim U/2$ which is not seen  in $\sigma_c(\omega)$.

\subsection{Low $T$ metallic regime }
\label{sec-lowT}
The lowest $T$ behavior with diverging $\kappa_\textrm{el}$ and $D_\textrm{Q,el}$,
can be described  within quasiparticle
picture. We start with the bubble formulas for both  $\kappa_\textrm{el}$ and $\sigma_\textrm{c}$ and with certain approximations (Appendix~\ref{app-quasi}) rewrite them in terms of the quasiparticle properties.
Using Eqs.~\ref{eq_kel} and \ref{eq_sc} for $\kappa_\textrm{el}$ and $\sigma_\textrm{c}$  with known approximations  \cite{kokalj13, krien19} 

\begin{eqnarray}
c_\textrm{el}&=&\frac{\pi^2}{3}\frac{g_0(\epsilon_\textrm{F})}{z} T,\label{eq_cel}\\
\chi_\textrm{c}&=& z g_0(\epsilon_\textrm{F}), \label{eq_chic}
\end{eqnarray}
one obtains via the Nernst-Einstein relation 
\begin{eqnarray}
D_\textrm{Q,el}\simeq z\frac{v^2_\textrm{0,F}}{-4\Sigma''(0)},\\
D_\textrm{c}\simeq z^{-1}\frac{v^2_\textrm{0,F}}{-4\Sigma''(0)}.
\end{eqnarray}

Here $g_0(\epsilon_\textrm{F})$ is a bare band density of states at Fermi energy, $v_\textrm{0,F}$ is bare Fermi velocity, $\Sigma''(0)$ is imaginary part of self energy at $\omega=0$ and $z=1/(1-\partial_\omega\Sigma'(\omega)|_{\omega=0})$ is the  quasiparticle renormalization. 

$D_\textrm{Q,el}$ is decreased from the bare non-renormalized value $D_0=v^2_\textrm{0,F}/[-4\Sigma''(0)]$ to $D_\textrm{Q,el}=z D_0$ with $z<1$ while $D_\textrm{c}=D_0/z$ is increased. Therefore in the Fermi liquid regime the two diffusion constants differ by a factor of $z^2$ and therefore easily by an order of magnitude. The question on renormalization effect for diffusion was posed already in Ref.~\onlinecite{pakhira15}. Note that only $D_\textrm{Q,el}$ can be expressed in terms of the quasiparticle velocity $v_\textrm{qp}=zv_\textrm{0,F}$ and life time $\tau_\textrm{qp}=[-2z\Sigma''(0)]^{-1}$ as $D_\textrm{Q,el}=v_\textrm{qp}^2 \tau_\textrm{qp}/2$, but not $D_\textrm{c}$.

\begin{figure}[ht!]
 \begin{center}
   \includegraphics[ width=0.99\columnwidth]{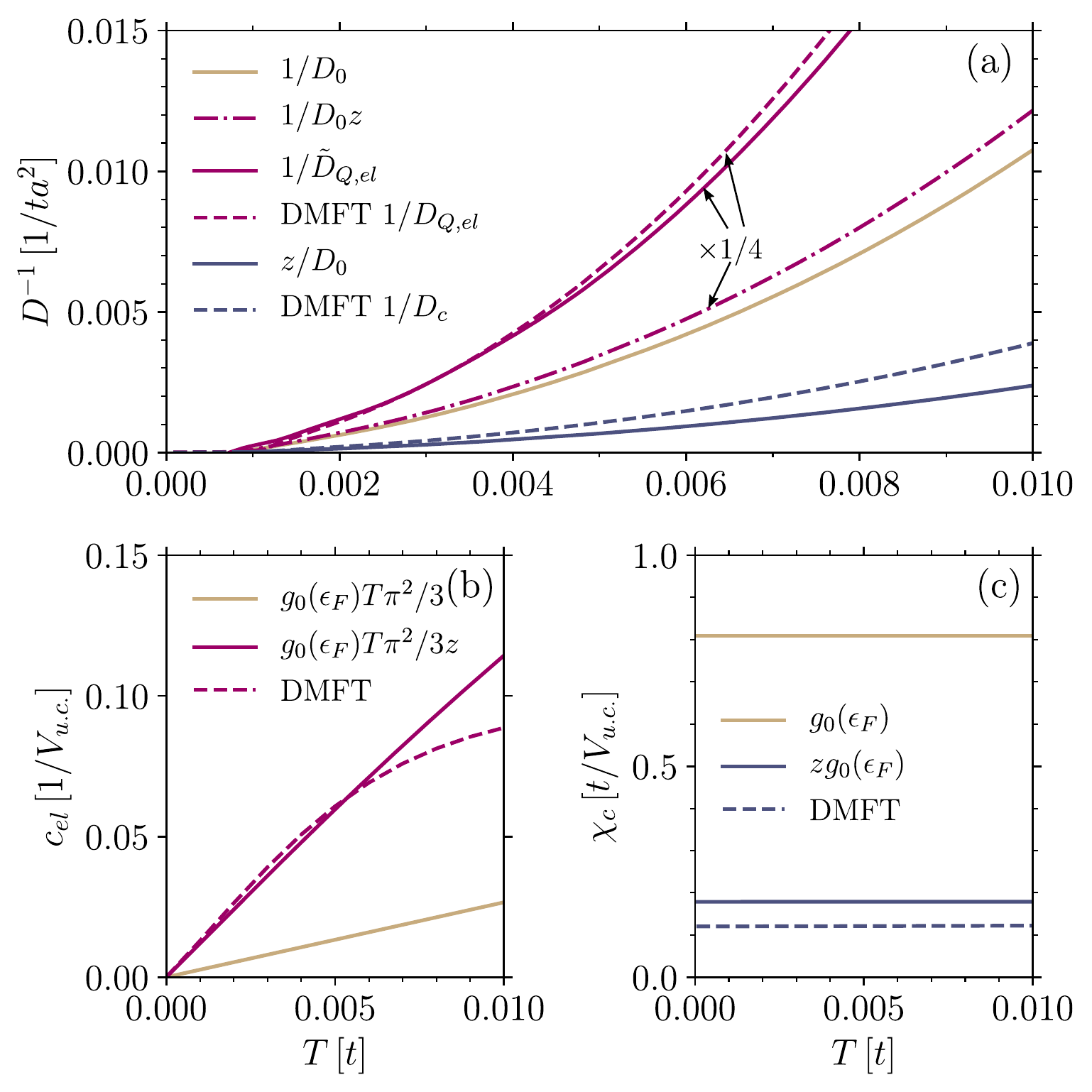}
\end{center}
  \caption{
DMFT results in the low-$T$ Fermi liquid regime compared with simplified expressions  (see main text) in terms of a quasi-particle properties.
(a) Comparison for diffusion constants.  All estimates for $D_\textrm{Q,el}$ (purple) are divided by 4 for compactness and clarity. (b) Comparison of $c_\textrm{el}$ with Sommerfeld expression with non-renormalized and renormalized density of states. (c) Comparisons of DMFT $\chi_\textrm{c}$ with $g_0(\epsilon_\textrm{F})$  and renormalized value $z g_0(\epsilon_\textrm{F})$.
$U=10t$, $p=0.15$, $z\approx 0.22$.
}
 \label{figdmft}
\end{figure}

We illustrate these considerations with the DMFT results \footnote{Since on a square lattice the Fermi velocity is not constant over the Fermi surface the $v_\textrm{0,F}^2$ is replaced with its averaged over the Fermi surface value $v_\textrm{0,F}^2=\int d^2k v_{0,k}^2 \delta(\epsilon_k-\epsilon_\textrm{F})/\int d^2k \delta(\epsilon_k-\epsilon_\textrm{F})$} shown in  Fig.~\ref{figdmft}.
$D_\textrm{c}$ agrees better with $D_0/z$ than with $D_0$ and $zD_0$ is
closer to DMFT $D_\textrm{Q,el}=\kappa_\textrm{el}/c_\textrm{el}$ than
$D_0$. Some mismatch persists in the latter, originating in our
oversimplification of assuming a constant $\Sigma''(\omega) \sim
\Sigma''(0)$ (see Appendix~\ref{app-vertex}). $\Sigma''(\omega)$
can have notable $\omega$ dependence (e.g., $\Sigma''(\omega) \propto
\omega^2$ in FL) and the factor $(-\frac{\partial n_\textrm{F}}{\partial
\omega})\omega^2$ in Eq.~\ref{eq_kel6} filters out finite frequencies.
Explicitly, frequencies at around $|\omega|\sim 3T$ are mainly included.
We therefore compare DMFT $D_\textrm{Q,el}$ also with $\tilde
D_\textrm{Q,el}=z v_\textrm{0,F}^2/[-4\Sigma''(\omega=3T)]$
\footnote{Since $(-\frac{\partial n_\textrm{F}}{\partial \omega})\omega^2$ mostly weights positive and negative frequencies $\omega\sim \pm 3T$ and in DMFT the $\Sigma''(\omega)$ is not particle-hole symmetric (even in $\omega$), the value of $\Sigma''(\omega=3T)$ should be understood as $\Sigma''(\omega=3T)=[\Sigma''(3T)+\Sigma''(-3T)]/2$}
and find good agreement. For completeness, Fig.~\ref{figdmft}(b,c) also show that $c_\textrm{el}$ and $\chi_\textrm{c}$ are well approximated with renormalized values given in Eqs.~\ref{eq_cel} and \ref{eq_chic}.

\section{The Wiedemann-Franz law}
\label{sec_wf}

\begin{figure}[ht!]
 \begin{center}
   \includegraphics[ width=0.99\columnwidth]{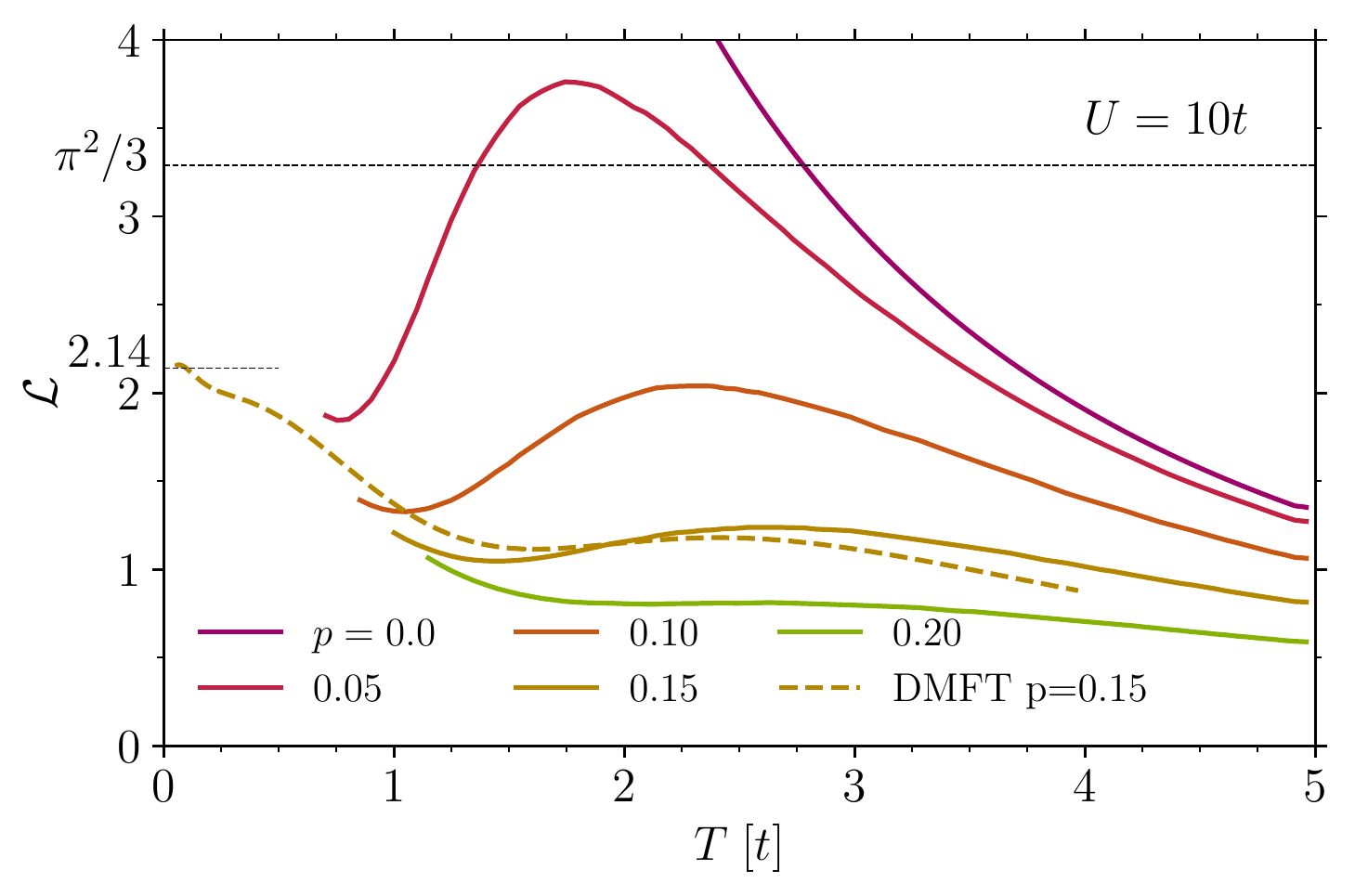}
\end{center}
 \caption{
Lorenz ratio vs. $T$ for several hole dopings calculated with FTLM. DMFT result for $p=0.15$ is shown. 
A low-$T$
pure Fermi liquid result $\mathcal L=2.14$ is also shown
\cite{pourovskii17}. 
}
 \label{fig3}
\end{figure}

The above discussion of different $T$ dependence of $D_\textrm{Q,el}$ and
$D_\textrm{c}$ suggests a possible violation of the Wiedemann-Franz (WF) law and
the deviation of the Lorenz ratio 
\begin{equation}
\mathcal L=\frac{\kappa_\textrm{el}}{T\sigma_\textrm{c}}
\label{eq_lorenz}
\end{equation}
from the Sommerfeld value $\pi^2/3$. In Fig.~\ref{fig3} we show the
calculated Lorenz ratio and observe in the whole calculated regime a clear
deviation from the Sommerfeld value. In addition, we find a strong $T$ dependence which is also non-monotonic for small $p$. 

We now discuss the various violations of the WF law. In the high $T$ limit this violation follows trivially as $L_{ij}\propto 1/T$, leading to
$\kappa_\textrm{el} \propto 1/T^2$ and $\sigma_\textrm{c} \propto 1/T$
\footnote{The same
dependencies are obtained by using constant diffusion constants and dependencies
$\chi_\textrm{c}\propto 1/T$ and $c_\textrm{el}\propto 1/T^2$.}.
This gives $\mathcal  L\propto
1/T^2 $ in the high $T$ limit \cite{perepelitsky16}, which is observed also in our results. The WF law is also violated for zero doping at low $T$, since
$\kappa_\textrm{el}$ has a nonzero spin
contribution while $\sigma_\textrm{c}$ is exponentially suppressed. $\mathcal L$ is therefore
expected to exponentially diverge as $T\to 0$, which is  seen in Fig.~\ref{fig3}. 

In the doped case at intermediate $T$, $\mathcal L$ is not well described by the Sommerfeld value either. One finds a non-monotonic $T$ dependence with a maximum. This maximum is most pronounced for low $p\sim 0.05$, where
$\mathcal L$ at higher $T$ is most similar to the undoped case. It is therefore possible to ascribe the maximum at $T\sim 2t$ to the spin contribution.
Even in the Fermi-liquid regime $\mathcal L\sim 2.14$ \cite{pourovskii17} 
due to $\kappa_\textrm{el}$ taking into account higher $\omega$ scattering rate, while $\sigma_\textrm{c}$ is
affected more by a $\omega \sim 0$ scattering rate. See Fig.~\ref{fig3} and section~\ref{sec-lowT}. 

$\mathcal{L}$ equals $\pi^2/3$ only for $\omega$-independent scattering rates (see Appendix~\ref{app-vertex}), which appears, e.g., in the elastic impurity dominated scattering and when vertex correction are negligible. This is demonstrated in Fig.~\ref{fig_ybco_L} with the AMFL model results at low $T$. The DQMC results for $\mathcal{L}$ in the Hubbard model are discussed in Ref.~\cite{wang22x}.

It is interesting to note that the DMFT result for $\mathcal L$ approximately agrees with the FTLM result as seen in Fig.~\ref{fig3} while the
conductivities $\kappa_\textrm{el}$ and $\sigma_\textrm{c}$ differ even by a factor
close to 2 as shown in the Appendix~\ref{app-vertex}. This suggest that in the
DMFT missing vertex corrections \cite{vucicevic19, vranic20} in
$\kappa_\textrm{el}$ and $\sigma_\textrm{c}$ almost
cancel when calculating $\mathcal L$.

\section{Discussion of experiments}
\label{sec_exper}
The purpose of this section is to discuss our results in terms of measurements on cuprates. The FTLM results do not reach sufficiently low temperatures for a direct comparison but the approximative extrapolations (spin-wave estimate in the undoped case and the DMFT in the doped case) do. For the doped case we also present the results of the phenonomenological anisotropic marginal Fermi liquid model and compare them to the measurements on overdoped cuprates in Appendix~\ref{app-amfl}.

\subsection{Mott-insulating LCO}

\begin{figure}[ht!]
 \begin{center}
   \includegraphics[ width=0.99\columnwidth]{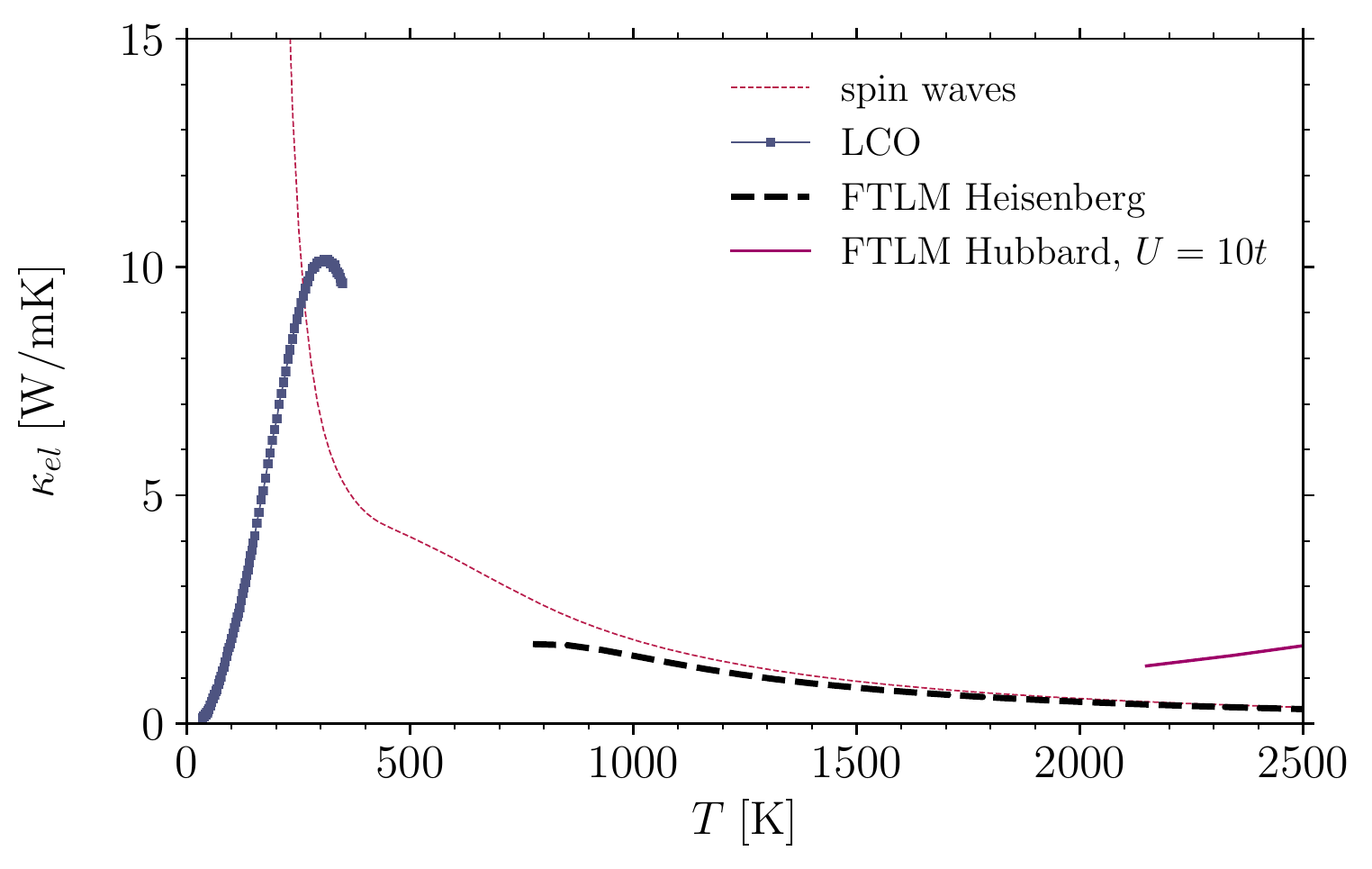}
\end{center}
 \caption{
Experimental $\kappa_\textrm{el}$ for  La$_2$CuO$_4$ taken from Ref.~\onlinecite{hess03} plotted next to theoretical estimates.
}
\label{fig_lco}
\end{figure}

Fig.~\ref{fig_lco} displays our results next to the measured  $\kappa_\textrm{el}$ \cite{hess03} in La$_2$CuO$_4$ (LCO), the parent Mott-insulating cuprate compound \footnote{For LCO we used parameters $J=1550\,$K, lattice constants $a_0=3.8\,$\AA\,and $c_0=13.2\,$\AA~\cite{hess03}\,with two CuO planes within $c_0$.}.
The measured data represent the magnonic (spin) contribution to $\kappa$ as the contribution from phonons was subtracted.  
A prominent feature in the measured data is the peak at $T\sim 300\,$K. It appears due to the saturation of $l$ at low $T$ (a finite impurity concentration)
and decreasing $c_\textrm{el}\to 0$ as $T\to 0$. On the high temperature side, measured $\kappa_\textrm{el}$ drops due to decreasing $l$.

Our FTLM data are reliable only at  higher $T$ and the results for the $U=10t$ Hubbard model still decrease with decreasing $T$. 
The Heisenberg model result at
lowest $T\sim 800\,$K is $\kappa_\textrm{el} \sim 2\,$W/mK. The measured $\kappa_\textrm{el}$ reaches $\sim 10\,$W/mK at $T \sim 400\,$K. To account for this increase, $l$ has to increase by a factor of $\sim 20$  (note that $c_\mathrm{el}$ is already decreasing). 
Such an increase is not observed in the DQMC Hubbard model results (see Fig.~\ref{fig_heis}a). 

Also instructive is to compare the data with the spin-wave estimate. Because in this we do not include the saturation of $l$ at low $T$ due to imperfections, the spin-wave $\kappa_\mathrm{el}$ diverges at low $T$. The growth is  moderate around $500\,$K but becomes more rapid at lower $T$ due to
a rapidly increasing $l$, e.g. $l \sim 50\,a$ at $300\,$K.
Interestingly, the spin-wave result is still lower than the experimental value at this $T$.

\subsection{The doped Mott insulator YBCO}

\begin{figure}[ht!]
 \begin{center}
   \includegraphics[ width=0.99\columnwidth]{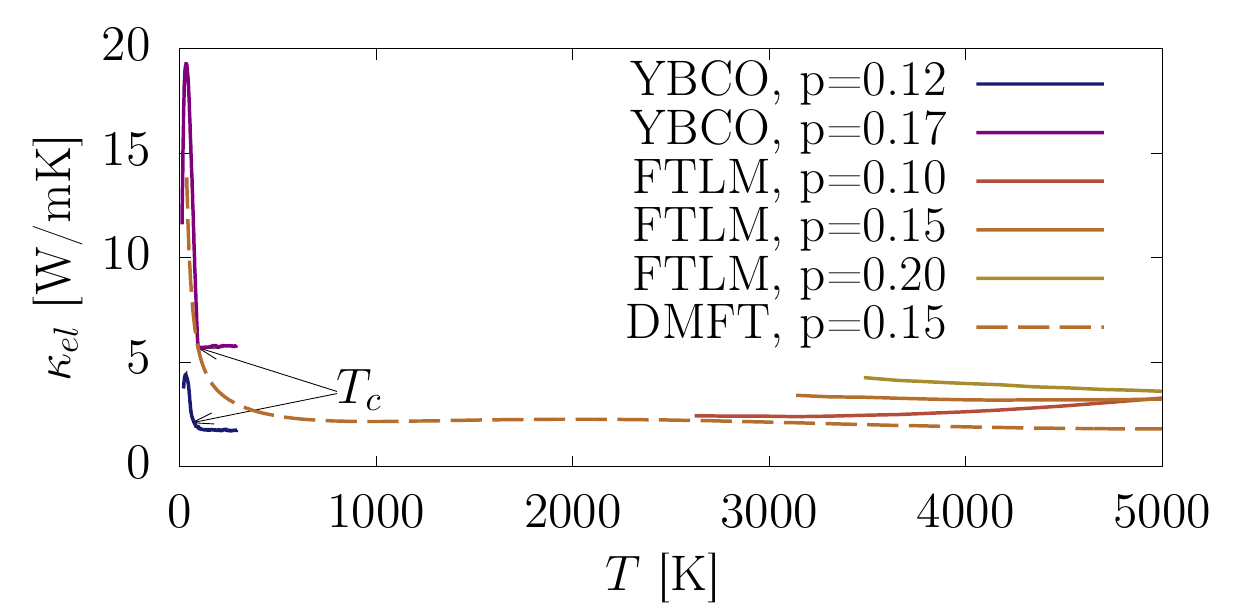}
\end{center}
 \caption{
Comparison of measured $\kappa_\textrm{el}$ for doped cuprate YBa$_2$Cu$_3$O$_{7-y}$ with Hubbard model results for U=10t. $\kappa_\textrm{el}$ are taken from Ref.~\onlinecite{takenaka97} and are obtained by subtracting the estimated $\kappa_\textrm{ph}$. The data are for samples with $y=6.68$ and $y=6.93$, which correspond to dopings $p=0.12$ and $p=0.17$, respectively. 
Above $T_\textrm{c}$ the experimental $\kappa_\textrm{el}$ shows weak $T$ dependence and similar magnitude as the DMFT and FTLM results at much higher $T\sim 3000\,$K.}
 \label{fig_ybco}
\end{figure}

In Fig.~\ref{fig_ybco} we compare our $\kappa_\textrm{el}$ results with the data for doped YBa$_2$Cu$_3$O$_{7-y}$ (YBCO) as reported in Ref.~\onlinecite{takenaka97}. Note that the measured data are available only at $T<300\,$K while the FTLM results are reliable only at $T>2500\,$K
\footnote{For YBCO we used unit cell parameters $a_0=3.82\,$\AA, $b_0=3.89\,$\AA \, and $c_0=11.68\,$\AA \, from Ref.~\onlinecite{bondarenko17,varshney11} together with $t=0.3$ eV and two CuO planes within $c_0$.}.
We plot also the DMFT result, which agrees with experimental data surprisingly well. 

The key aspect of the data is that the values of $\kappa_\textrm{el}$ in FTLM at $3000\,$K are quite close to the measured ones at $300\,$K and that DMFT indicates a $T$ independent $\kappa_\textrm{el}$. We expect that the actual $\kappa_\textrm{el}$ is higher than the DMFT one.
The FTLM result on $\kappa_\textrm{el}$ for doping
$p=0.15$ shows a rather weak $T$ dependence in Fig.~\ref{fig_ybco}, which is reminiscent of
experimental behavior \cite{takenaka97,minami03}, albeit at significantly higher $T$. 

It is worth mentioning that linear-in-$T$ $c_\textrm{el}$ and linear-in-$T$ electrical resistivity due to scattering rate (e.g., $D\propto 1/T$ in a bad or strange metal) can lead to $T$-independent or close to constant $\kappa_\textrm{el}$, at least for Lorenz ratio with weak $T$ dependence.

We note that the measured data show a strong increase below $T_c$. This is a superconducting effect and it has been suggested that it appears due to the increased coherence of non-superconducting electrons \cite{waske07}. The low-T increase of DMFT results is not related to superconductivity, but originates in the increased coherence and longer $l$. However, such increase could be suppressed or restricted to lower $T$ by spin, charge or order parameter phase fluctuations, which are not included in DMFT. These could explain the weak $T$ dependence measured in the normal state.

Let us note in passing that $D_\textrm{Q}$ was measured for several cuprates \cite{zhang17, zhang19} to have typical values of 0.02\,cm$^2$/s at high $T$ ($\sim 300 - 600\,$K). Taking the estimated lower bound on electronic thermal diffusion $D_\textrm{Q,el,MIR}=ta^2 \sim 0.7\,$cm$^2$/s, it is evident that the measured $D_\textrm{Q}$ is smaller by an order of magnitude. This is due to the heat transfer being dominated by phonons, which have orders of magnitude smaller velocity than electrons. 

It is also interesting to note that $\kappa_\textrm{el}$ in the Mott insulator~\cite{hess03} has values around $10\,$W/mK at $300\,$K (Fig.~\ref{fig_lco}), which is an order of magnitude larger than in the low doping metallic phase \cite{yu92,takenaka97,zhang17}, where $\kappa\approx 1\,$W/mK (Fig.~\ref{fig_ybco}).
The magnonic part of $c _\textrm{el}$ ($\propto T^2$) is smaller than in the doped case ($\propto T$, see  also lowest $T$ results in Fig.~\ref{fig3}b) and the spin-wave velocity ($v\sim Ja$) is smaller than the particle velocity ($v\sim ta$). 
This indicates that quasiparticles have orders of magnitude smaller $l$ than spin-waves and that their transport is at low doping significantly less coherent.

This is also in accord with strongly underdoped YBCO ($y=6.34$) being taken in Ref.~\onlinecite{takenaka97} as a case with smallest $\kappa_\textrm{el}$. We discuss the experimental Lorenz ratio for YBCO in Appendix~\ref{app-amfl-lorenz}, where we compare it to the AMFL model results and discuss the relation of its temperature dependence to the frequency dependence of the scattering rate.

\section{Conclusions}
\label{sec_concl}

We studied $\kappa_\textrm{el}$ and $D_\textrm{Q,el}$ with numerical calculations of square lattice Hubbard and Heisenberg models and further with phenomenological models. 

In the  Mott-insulating phase $\kappa_\textrm{el}$ is nonmonotonic and has three features. It has a high-T peak related to a peak in $c_\mathrm{el}$ due to charge excitations. $c_\mathrm{el}$ has another peak at $T\sim 0.6 J$ and drops at lower $T$ due to the quenching of the spin-entropy. This happens at $T$ below that accessible to us in the numerical calculations of transport. However in our spin-wave phenomenological estimate the dependence of $c_\mathrm{el}$ at $T\sim J$ does not lead to a peak in $\kappa_\mathrm{el}$, only a shoulder, while the situation in the Hubbard and Heisenberg models remains unsettled. At even lower $T$, $\kappa_\mathrm{el}$ peaks (diverges in a pure model) due to increased $l$.
The Hubbard model $\kappa_\textrm{el}$ and $D_\textrm{Q,el}$ approach the Heisenberg model results with decreasing $T$. At lower $T<J$, the difference between the Heisenberg model results and DQMC Hubbard results deserves further study. 

We introduced a MIR value for thermal conductivity and found that at higher $T$, $\kappa_\textrm{el}$ is below it, thus violating the na\"ive bound. Conversely, the calculated thermal diffusion $D_\textrm{Q,el}$ has values that correspond to $l\gtrsim a$ (except for smallest dopings that can be perhaps understood in terms of suppressed velocity). The thermal transport thus behaves analogously to the charge transport in this respect. The analogy is however incomplete. 
We compared the temperature dependences of $D_\textrm{Q,el}$ and $D_\textrm{c}$ and found they behave differently: in the intermediate temperature regime $D_\textrm{Q,el}$ has a more incoherent behavior and less temperature dependence than $D_\textrm{c}$. In the well defined quasiparticle regime $D_\textrm{Q,el}$ and $D_\textrm{c}$ behave differently upon renormalization and differ by a factor of $z^2$. The renormalization in cuprates is typically $1/z=3-4$, thus $D_\textrm{Q,el}$ and $D_\textrm{c}$ can differ by an order of magnitude. The discussion of experimental data in terms of effective velocities $v$ and $l$ and the discussions of diffusion bounds \cite{hartnoll15,zhang17}
should take these distinctions and the influence of renormalization properly into account. 

We find that $\mathcal L$ 
depends strongly on $T$ and that the WF law is typically violated 
with $\mathcal L$ being either larger or smaller than the Sommerfeld value $\pi^2/3$ in several regimes even by a factor of $\sim 2$.
All these deviations bring a clear message: the usual practice of estimating $\kappa_\textrm{el}$ using the WF law is problematic.

The temperatures of our FTLM simulations are well above the experimental ones, yet it is interesting to note the different status of the Mott-insulating case where the much larger experimental thermal conductivity 
points to a very rapid growth of $l$ with lowering $T$. Conversely, in the doped case the measured $\kappa_\textrm{el}$ has similar values as the Hubbard model result at much higher temperatures,
suggesting that $\kappa_\textrm{el}$ is only weakly temperature dependent in between.

We explore lower temperatures within the DMFT approximation and indeed find such behavior. 
We also present results obtained with a phenomenological AMFL model and give predictions for  $\kappa_\textrm{el}$ and the Lorenz ratio for the overdoped cuprates in the temperature regime of experiments (see   Appendix~\ref{app-amfl}).

For data availability, see Ref.~\onlinecite{notedata}.

\section*{Acknowledgments}
 This work was supported by the
Slovenian Research Agency (ARRS) under Program No. P1-0044.
Part of computation was performed on the supercomputer Vega at the Institute of Information Science (IZUM) in Maribor.

 \appendix{

\section{Details of the FTLM calculations}
\label{app-ftlm}

The transport coefficients are defined via

\begin{align}
    j_\textrm{n}=-L_{11}\nabla \mu-L_{12}\nabla T/T,\\
    j_\textrm{Q}=-L_{21}\nabla\mu-L_{22}\nabla T/T.
\end{align}
\noindent
Here $L_{11}=\sigma_\textrm{n,n}$, $L_{12}=\sigma_\textrm{n,Q}=L_{21}$, $L_{22}=\sigma_\textrm{Q,Q}$ are conductivities which are obtained by zero frequency limit ($\omega=0$) of the generalized conductivities at finite frequency $\omega$ \cite{shastry09}. These are calculated via the generalized susceptibility

\begin{align}
    \sigma_\textrm{A,B}(\omega)&=  \frac{\chi_\textrm{A,B}''(\omega)}{\omega},\\
    \chi_\textrm{A,B}(\omega)&=\frac{i}{N V_{u.c.}}\int_0^{\infty} dt e^{i\omega t}\langle [\hat{J}_\textrm{A}(t),\hat{J}_\textrm{B}(0)]\rangle.
\end{align}
\noindent
$\hat{J}_A$ and $\hat{J}_B$ are the current operators, associated with (conserved) quantities $\hat{A}$ and $\hat{B}$, which can be derived from polarisation

\begin{align}
    &\hat{P}_{A}=\sum_j x_j\hat{A}_j, &\hat{J}_\textrm{A}=\frac{d\hat{P}_\textrm{A}}{dt}=i[\hat{H},\hat{P}_\textrm{A}].
\end{align}

For the single band Hubbard model, the operators for particle and heat currents in $x$ direction are given by

\begin{align}
    \hat{J}_n=&-it\sum_{j,\sigma,\delta} R_\delta^x c^{\dagger}_{j+\delta,\sigma}c_{j,\sigma},\\
    \hat{J}_\textrm{E}=&-\frac{it^2}{2}\sum_{j,\sigma,\delta,\delta '} R_{\delta\delta'}^x c^{\dagger}_{j+\delta+\delta',\sigma}c_{j,\sigma}\nonumber 
            \\&+\frac{itU}{2}\sum_{j,\sigma,\delta}R_\delta^x c^{\dagger}_{j+\delta,\sigma}c_{j,\sigma}(n_{j+\delta,\bar \sigma}+n_{j,\bar \sigma}),\\
    \hat{J}_\textrm{Q}=&\hat{J}_\textrm{E}-\mu\hat{J}_n,
\end{align}
\noindent

where $R_{\delta}^x=x_{j+\delta}-x_j$ and $R_{\delta\delta'}^x=x_{j+\delta+\delta'}-x_j$.

We also consider the 2D Heisenberg model with nearest neighbor interaction
\begin{equation}
\hat{H}\! =\! J \sum_{\langle ij \rangle }  \hat{{\bf S}}_i \cdot  \hat{{\bf S}}_j\! =\! \sum_i \hat{H}_i, \qquad 
\hat{H}_i\! =\! J \hat{{\bf S}}_i \cdot  (\hat{{\bf S}}_{i+1_x} + \hat{{\bf S}}_{i+1_y} ),
\end{equation}
where the sum runs over sites $i$ on the square lattice.
The energy current operator for the Heisenberg model reads
\begin{gather}
\hat{J}_\textrm{E} = -i J^2\sum_i [ \hat{O}_{i-1_x,i,i+1_x} + \hat{O}_{i-1_x,i,i+1_y} \nonumber\\ + \hat{O}_{i-1_x,i,i-1_y} ], \\
\hat{O}_{ijl} = \frac{1}{2} [ \hat{S}_i^z \hat{S}^\pm_{jl} +\hat{S}_j^z \hat{S}^\pm_{li} + \hat{S}_l^z \hat{S}^\pm_{ij} ],\\
\hat{S}^\pm_{jl} = \hat{S}^+_j  \hat{S}^-_l - \hat{S}^+_l  \hat{S}^-_j . 
\end{gather}

The spectra of dynamical quantities on finite clusters consist of $\delta$ functions which we broaden using a Gaussian kernel. Finite size effects at low-$T$ manifest also as a growing contribution of the $\delta$ function at $\omega=0$. We can thus use this as a criterion to estimate the lowest temperature at which we can trust our FTLM results and set the maximum acceptable fraction of the total spectra contained in $\delta(0)$ to $\lesssim 0.5\%$.

At the particle-hole symmetric point at half filling in the Hubbard model, $L_{12}$ vanishes. Thus, $\sigma_\textrm{n,E}=\mu\sigma_\textrm{n,n}$ and $\kappa_\textrm{el}=(\sigma_\textrm{E,E}-\mu^2\sigma_\textrm{n,n})/T$ with $\mu=U/2$. Where $\sigma_\textrm{n,n}$  has a charge gap and is exponentially suppressed (at temperatures in the Heisenberg regime), only $\sigma_\textrm{E,E}$ contributes to $\kappa_\mathrm{el}$.

\section{Cluster size and shape dependence for the Heisenberg model}
\label{app-cluster}

\begin{figure}[ht]
 \begin{center}
   \includegraphics[ width=0.99\columnwidth]{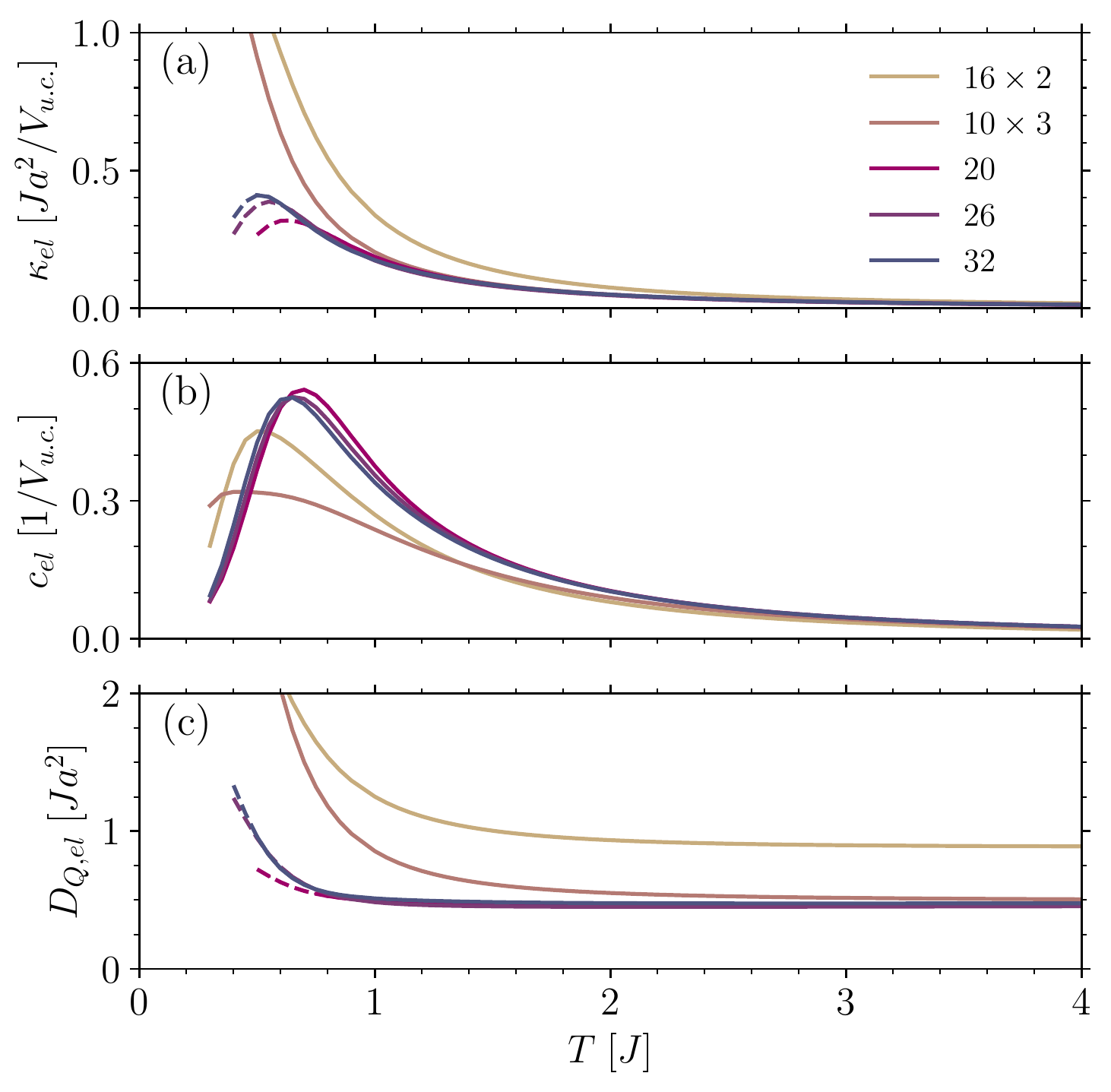}
\end{center}
 \caption{
Heat conductivity, specific heat and heat diffusion constant for different geometries in the Heisenberg model. We consider 2 and 3 rung ladders (labeled $16\times 2$ and $10\times 3$, respectively) with periodic boundary conditions also in the transverse direction. 2d lattices with 20, 26 and 32 sites are tilted to allow unfrustrated antiferromagnetic correlations. Where finite-size effects are significant, we show the data dashed. 
}
\label{fig-app-b}
\end{figure}

Here we discuss in  more detail the Heisenberg model results and in particular the possibility of $\kappa_\textrm{el}$ having a peak at $T\sim 0.6 J$ where $c_\textrm{el}$ shows a peak. In Fig.~\ref{fig-app-b} we show $\kappa_\textrm{el}$, $c_\textrm{el}$ and $D_\textrm{Q,el}$ for several cluster sizes and shapes. The results for $\kappa_\textrm{el}$ for square lattices show clear finite size effect at  $T \lesssim  0.6J$. Although the results for $\kappa_\textrm{el}$ on square clusters do show a maximum corresponding to the maximum in $c_\textrm{el}$, the system size dependence at the maximum is still considerable and thus we cannot exclude the absence of the maximum in the thermodynamic limit. I.e., the mean free path becomes too large at low $T$ making the finite-size effect too large to discriminate between a peak or a shoulder in $\kappa_\textrm{el}$.
For comparison we also investigated $\kappa_\textrm{el}$ for 2 leg and 3 leg ladders. These show a much stronger increase of $D_\textrm{Q,el}$ with decreasing $T$ and therefore much more coherent behavior for $T\lesssim J$. Even in these long systems, we cannot conclusively determine the existence of a peak in $\kappa_\textrm{el}$ at $T$ where $c_\textrm{el}$ has the peak, again due to finite size effect and long mean free paths. However, due to the strong increase of the mean free path with decreasing $T$, the absence of the peak and appearance of a shoulder is more likely. The ``spin wave" result in Fig.~\ref{fig1} in the main text supports this scenario. 
It is also worth mentioning that the materials with 2 leg ladders are one of the best thermal conductors \cite{hess07} despite having a spin gap of $\Delta_\textrm{s} \sim 0.5J$} \cite{dagotto99}.

 \section{Frequency spectra}
\label{app-freq}
 
\begin{figure}[ht]
 \begin{center}
   \includegraphics[ width=0.99\columnwidth]{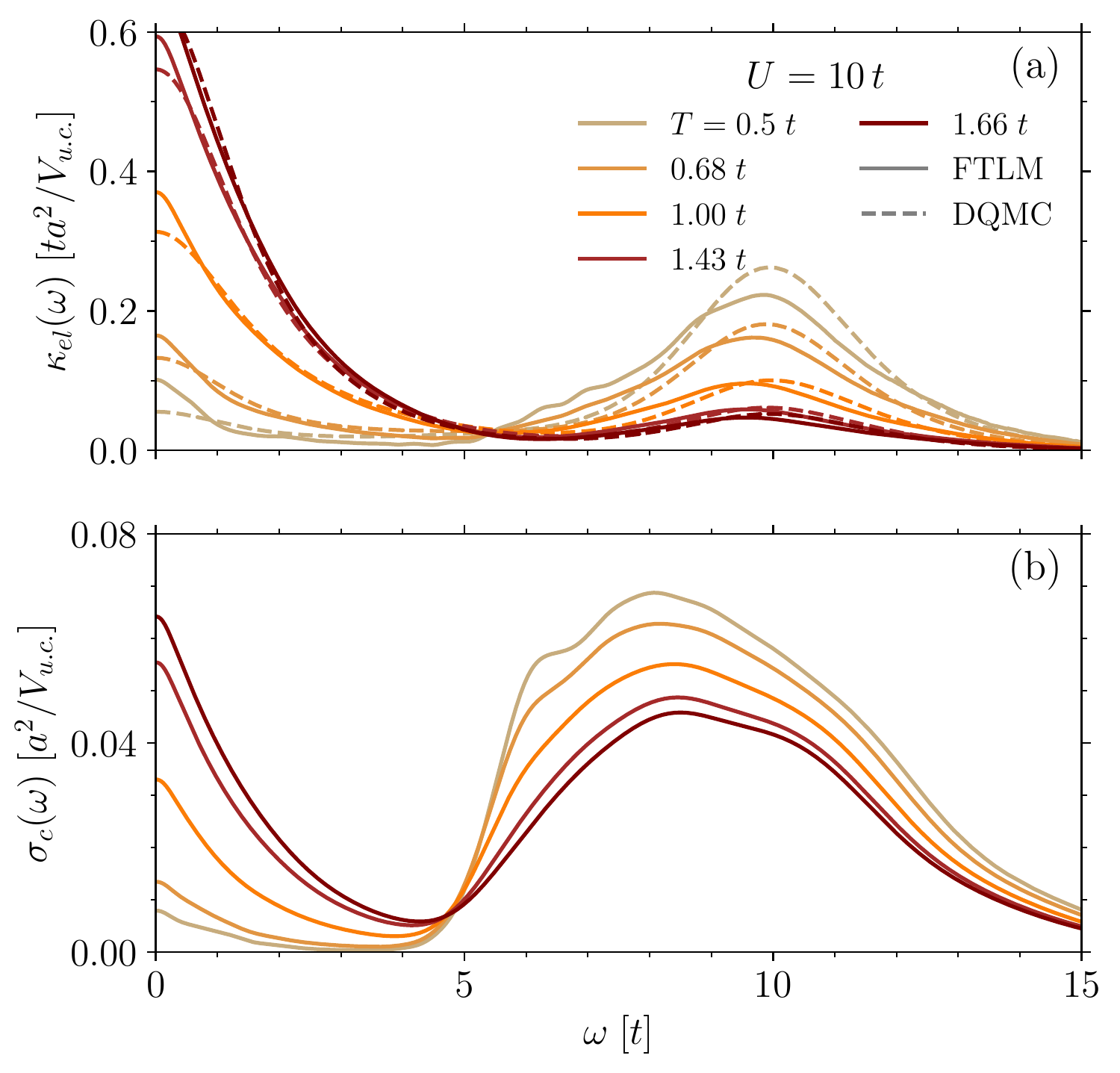}
\end{center}
 \caption{
The spectra $\kappa_\textrm{el}(\omega)$ and $\sigma(\omega)$ at different temperatures for $U=10\,t$ and at half filling. DQMC results
are taken from Ref.~\onlinecite{wang21}.
}
 \label{fig-app-a}
\end{figure}

 The temperature evolution of optical spectra of  $\kappa_\textrm{el}(\omega)$ and $\sigma_\textrm{c}(\omega)$
 is shown on Fig.~\ref{fig-app-a}. One sees a Drude-like peak at low $\omega$
 and a Hubbard satellite peak at $\omega\sim U$ with the two separated by a  gap, which is 
 most pronounced in $\sigma_\textrm{c}(\omega)$ at low $T$. $\sigma_\textrm{c}(\omega)$ clearly exhibits a gap 
 and the weight of the Drude peak raises as $T$ is increased. In the high 
 $T$ limit, the weight of this peak is comparable to the satellite peak. In contrast, 
 $\kappa_\textrm{el}(\omega)$ does not show a clear gap or gap edge. In addition, the satellite peak is significantly smaller than the Drude peak at 
 elevated $T$. This decrease in the relative weights of the two peaks originates in the 
 difference of frequency dependence of $\sigma_\textrm{E,E}(\omega)$ and $\sigma_\textrm{c}(\omega)$.
 
 FTLM results agree  well with those from DQMC in particular at higher $T$. FTLM has a sharper 
 Drude peak which deviates from a simple Lorentz shape. Because of this, FTLM also gives somewhat higher dc values for $\kappa_\textrm{el}$ than DQMC. 
 
\section{Vertex corrections}
\label{app-vertex}

\begin{figure}[ht]
 \begin{center}
   \includegraphics[width=0.99\columnwidth]{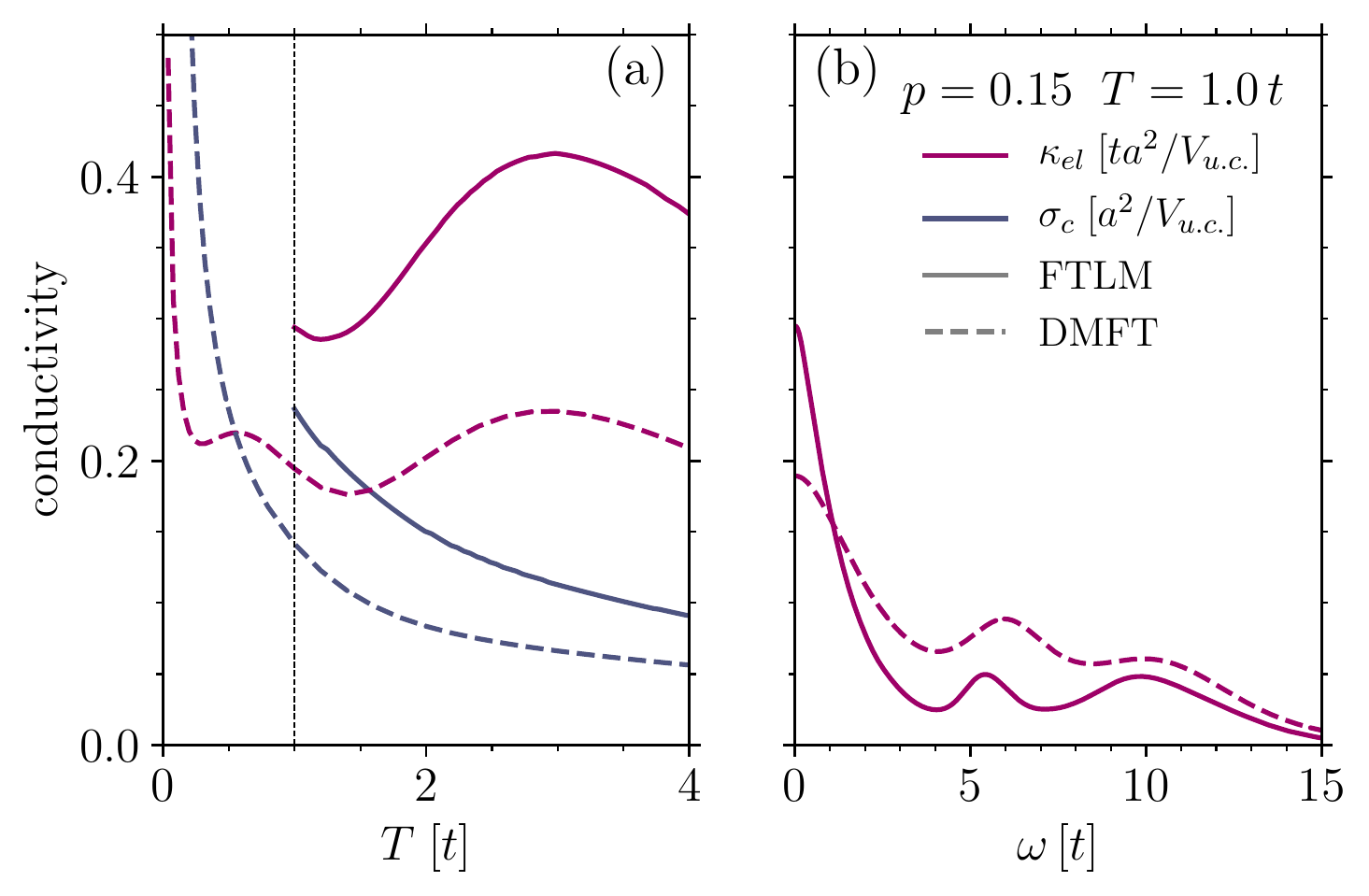}
\end{center}
 \caption{
Comparison of FTLM (solid) and DMFT (dashed lines) results for charge and heat conductivities
as a function of temperature (a) and frequency (b) for the Hubbard model with $U=10t$ and doping $p=0.15$. The frequency dependence is shown for the lowest considered temperature $T= t$ and denoted with a dashed line in 
panel (a).
}
 \label{fig-app-d}
\end{figure}

Recent investigations \cite{vucicevic19,vranic20} have tested the approximations of using local a self-energy and neglecting vertex corrections for calculation of dc charge conductivity $\sigma_\textrm{c}$. It was realized that the vertex corrections are substantial in the whole $T$ regime.
However, $\kappa_\textrm{el}$ was not considered and we show in Fig.~\ref{fig-app-d} $\kappa_\textrm{el}$ as a function of temperature and frequency as calculated with FTLM and DMFT. DMFT approximates self energy with a local quantity and neglects the vertex corrections.
For both, $\sigma_\textrm{c}$ and $\kappa_\textrm{el}$, DMFT  underestimates the DC conductivity as is shown in Fig.~\ref{fig-app-d}a,
with the difference $\sim $50\% for both $\sigma_\textrm{c}$ and $\kappa_\textrm{el}$. From this, one can say that the vertex corrections are substantial also for $\kappa_\textrm{el}$ and that they have similar magnitude than for $\sigma_\textrm{c}$. Regarding the frequency dependence (see Fig.~\ref{fig-app-d}b), the inclusion of vertex corrections makes the low-$\omega$ (Drude) peak narrower. 

However, both calculations point to the existence of a third peak in $\kappa_\textrm{el}(\omega)$ at $\omega\sim U/2$. Such a peak appears below $T\sim 1.7t$. We attribute this to the frequency dependence of $L_{12}(\omega)$, which has a zero at $\omega\sim U/2$. Thus, the frequency dependent Seebeck coefficient vanishes and heat transport is no longer suppressed by the term $L_{12}^2/(T L_{11})$ in Eq.~\ref{eq_kappa}. This leads to the peak in $\kappa_\textrm{el}(\omega \sim U/2)$. Such term and therefore the suppression can be traced back to the boundary condition of $j_\textrm{n}=0$ leading to a buildup of charges on sample edges.

\section{Diffusion constant with particle properties in the quasiparticle regime}
\label{app-quasi}

Here we start with the bubble expressions for thermal conductivity $\kappa_\textrm{el}$ and charge conductivity $\sigma_\textrm{c}$,
\begin{eqnarray}
\kappa_\textrm{el}&=&\frac{1}{T}\!\!\int\!\! d\omega (-\frac{\partial n_\textrm{F}}{\partial \omega})\omega^2 \frac{2\pi}{V}\sum_k v^2_{0,k,x} [A(k,\omega)]^2,\label{eq_bubble_kappa} \\
\sigma_\textrm{c}&=&\int\!\! d\omega (-\frac{\partial n_\textrm{F}}{\partial \omega}) \frac{2\pi}{V}\sum_k v^2_{0,k,x} [A(k,\omega)]^2,\label{eq_bubble_sigma} 
\end{eqnarray}
and rewrite them after some manipulation and use of approximations in terms of a quasiparticle properties, e.g, velocity, self-energy, density and renormalization.
$n_\textrm{F}$ is the $\omega$ dependent Fermi function, $v_{0,k,x}$ is the $x$ component of bare band velocity at $k$ point in the Brillouin zone, and $A(k,\omega)$ is the spectral function. 

Using the isotopic property of the square lattice one can replace  $v^2_{0,k,x} \to  v^2_{0,k}/2$. Further, one can use the velocity (or average velocity) at a certain energy to replace $ v^2_{0,k}\to  v^2_{\epsilon}$ and if one neglects $k$ dependence of self energy so that $A(k,\omega)$ is just a function of $\epsilon=\epsilon_k-\mu$ (and $\omega)$ one can write
\begin{equation}
\frac{2\pi}{V} \sum_k \frac{v^2_{0,k}}{2} [A(k,\omega)]^2=\pi\int d\epsilon g_0(\epsilon)\frac{v^2_{\epsilon}}{2} [A(\epsilon,\omega)]^2.
\end{equation}
Here noninteracting density of states  $g_0(\epsilon)=(2/V)\sum_k\delta(\epsilon_k-\mu-\epsilon)$
is introduced. Assuming that $g_0(\epsilon)$ and $v^2_\epsilon$ can be in the low $T$ regime replaced by constants and with the values at the Fermi energy, namely  $g_0(\epsilon_\textrm{F})$ and $v^2_\textrm{0,F}$, than the integral over $\epsilon$ can be performed. 
\begin{equation}
 \int d\epsilon  [A(\epsilon,\omega)]^2=\frac{1}{-2\pi\Sigma''(\omega)}.
\end{equation}
This leads to the following expressions for the conductivities.
\begin{eqnarray}
\kappa_\textrm{el}&=&\frac{1}{T}\int d\omega (-\frac{\partial n_\textrm{F}}{\partial \omega})\omega^2
g_0(\epsilon_\textrm{F})\frac{v^2_\textrm{0,F}}{-4\Sigma''(\omega)},\label{eq_kel6}\\
\sigma_\textrm{c}&=&\int d\omega (-\frac{\partial n_\textrm{F}}{\partial \omega}) g_0(\epsilon_\textrm{F})\frac{v^2_\textrm{0,F}}{-4\Sigma''(\omega)}.\label{eq_sc7}
\end{eqnarray}
Here $v_\textrm{0,F}$ is the bare Fermi velocity. 
At low $T$ the Fermi function derivative filters out only  low $\omega$  and we can roughly approximate the imaginary part of self energy with a constant, $\Sigma''(\omega)\sim \Sigma''(0)$. The  integral over $\omega$ can than be performed, which leads to
\begin{eqnarray}
\kappa_\textrm{el}&=&\frac{\pi^2}{3}
g_0(\epsilon_\textrm{F}) T \frac{v^2_\textrm{0,F}}{-4\Sigma''(0)},\label{eq_kel}\\
\sigma_\textrm{c}&=& g_0(\epsilon_\textrm{F})\frac{v^2_\textrm{0,F}}{-4\Sigma''(0)}.\label{eq_sc}
\end{eqnarray}
This nicely demonstrates that $\kappa_\textrm{el}$ and $\sigma_\textrm{c}$ are given in terms of non-renormalized quantities like bare band density of states at Fermi energy  $g_0(\epsilon_\textrm{F})$, bare Fermi velocity $v_\textrm{0,F}$ and bare particle scattering rate $\Gamma=-2\Sigma''(0)$.
Here $\Sigma''(0)$ is imaginary part of self energy at $\omega=0$. 

Above equations also indicate which terms are related to static thermodynamic properties like $c_\textrm{el}$ and $\chi_\textrm{c}$ and which to the diffusion constants. Note however, that these expression do not depend on the renormalization $z$, while $c_\textrm{el}$ and $\chi_\textrm{c}$ do. See main text for further discussion. 

By calculating the Lorenz ratio $\mathcal L$ (Eq.~\ref{eq_lorenz}) from Eqs.~\ref{eq_kel} and \ref{eq_sc}  one obtains the Sommerfeld value, $\mathcal L=\pi^2/3$. Note however, that here the $\omega$ dependence of $\Sigma''(\omega)$ was neglected and as discussed in the main text, $\omega$ dependence of $\Sigma''(\omega)$ effectively changes $\Sigma''(0)$ to $\Sigma''(\omega=3T)$ in expression for $\kappa_\textrm{el}$ (Eq.~\ref{eq_kel}). This moves $\mathcal L$ away from $\pi^2/3$ and therefore $\mathcal L=\pi^2/3$ only for the $\omega$-independent scattering rate.

\section{Anisotropic marginal Fermi liquid model for overdoped cuprates}
\label{app-amfl}

\subsection{Thermal conductivity of Tl2201}
\label{app-amfl-kappa}

\begin{figure}[ht!]
 \begin{center}
   \includegraphics[ width=0.99\columnwidth]{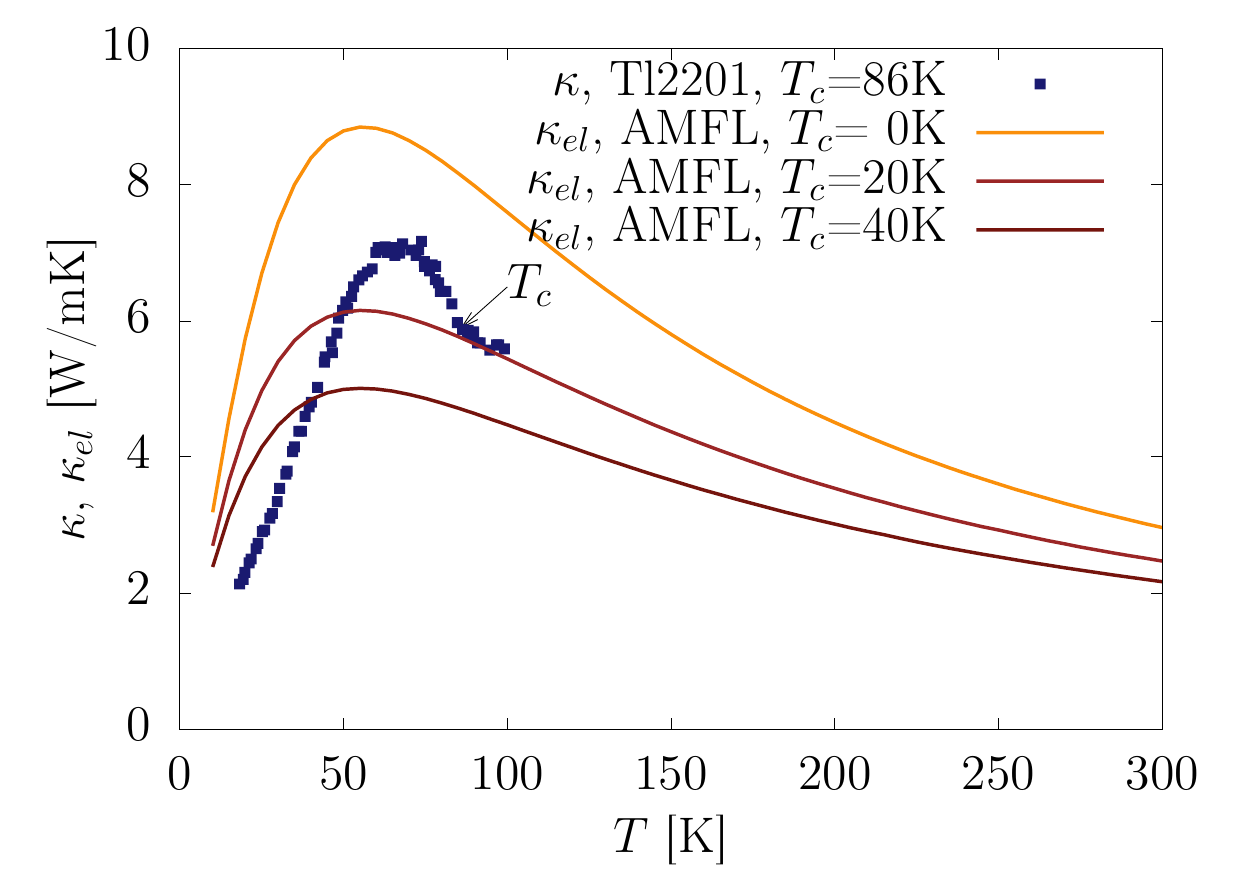}
\end{center}
 \caption{
Comparison of measured total $\kappa$ for Tl2201 with $T_\textrm{c}=86\,$K \cite{yu94,yu92}, with calculated electronic $\kappa_\textrm{el}$ within AMFL model \cite{kokalj11,kokalj12} for $T_\textrm{c}=0\,$K, $20\,$K and $40\,$K on the highly overdoped side. }
 \label{fig_tl2201}
\end{figure}

It is instructive to estimate also the electronic contribution to $\kappa$  within the phenomenological approach in the measured temperature regime.
For that we employ the anisotropic marginal Fermi liquid (AMFL) model \cite{kokalj11,kokalj12}, which was devised from the angle-dependent magnetoresistance experiments \cite{abdel06, abdel07, french09} on  overdoped Tl$_2$Ba$_2$CuO$_{6+\delta}$ (Tl2201) and quantitatively describes the elastic impurity scattering, the isotropic Fermi-liquid-like scattering, and the anisotropic marginal-Fermi-liquid-like scattering. The AMFL model captures the specific heat, mass renormalization and ARPES experiments \cite{kokalj11}, as well as resistivity, optical conductivity, magnetoresistance and Hall coefficient \cite{kokalj12} in Tl2201. 
We use the same parameters for the AMFL model as in Ref.~\onlinecite{kokalj12} and show results for $\kappa_\textrm{el}$ in 
 Fig.~\ref{fig_tl2201}. The results are obtained within the bubble approximation (Eq.~\ref{eq_bubble_kappa}) and are calculated for several dopings on the overdoped side, indicated by the values of the corresponding superconducting transition temperature $T_\textrm{c}$ ($0\,$K, $20\,$K and $40\,$K). The scattering rate becomes larger and more linear in $T$ and $\omega$ as one reduces doping towards the optimal doping. Therefore, $\kappa_\textrm{el}$ becomes smaller and also more constant in $T$ (since $c_\textrm{el}\propto T$ and $D_\textrm{Q,el}$ approaches $1/T$ behavior).
The AMFL results also show a maximum at $T\sim 50\,$K since $D_\textrm{Q,el}$ saturates at the lowest $T$ due to elastic impurity scattering while $c_\textrm{el}$ is decreasing with decreasing $T$. This is an example of the low-$T$ peak in $\kappa_\textrm{el}$ due to saturation of $l$. Note that no superconducting effects are captured within the AMFL model. The results are compared to the total $\kappa$ as measured in Tl2201 \cite{yu94,yu96}. Unfortunately, experimental data are available only close to optimal doping with $T_\textrm{c}=86\,$K (where AMFL is less valid) and report only total $\kappa$. It would be interesting to test the AMFL predictions for $\kappa_\mathrm{el}$ of Fig.~\ref{fig_tl2201} in the highly overdoped regime.

\subsection{The Lorenz ratio in YBCO}
\label{app-amfl-lorenz}

\begin{figure}[ht!]
 \begin{center}
   \includegraphics[ width=0.99\columnwidth]{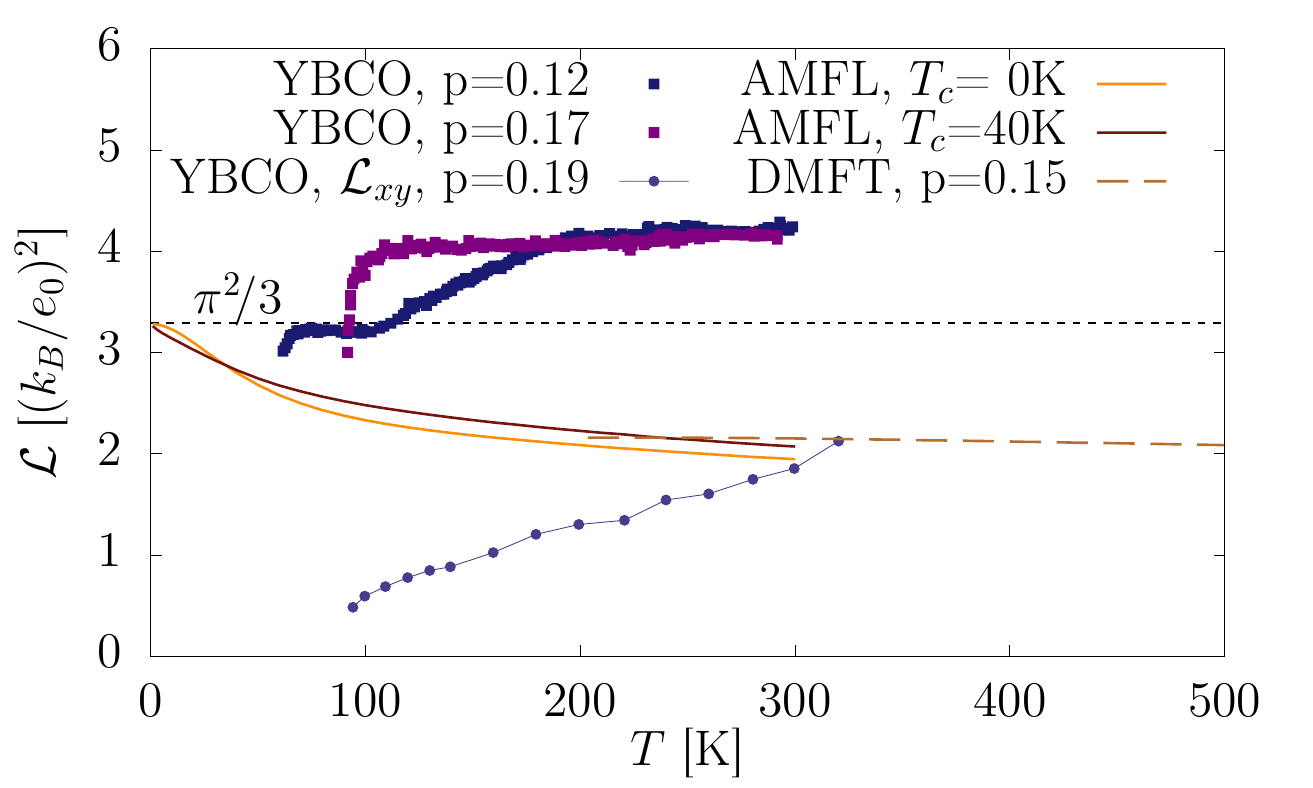}
\end{center}
 \caption{
Comparison of several measured data for Lorenz ratio $\mathcal{L}$ for YBCO (taken from Ref.~\onlinecite{takenaka97,zhang00}),  with theoretical estimates from an AMFL model for overdoped cuprates \cite{kokalj11,kokalj12}  and with DMFT calculation.  
}
 \label{fig_ybco_L}
\end{figure}

In Fig.~\ref{fig_ybco_L} we show theoretical estimates for the Lorenz ratio from the AMFL model and the DMFT result. The AMFL results tend to the Sommerfeld value as $T\to 0$ due to elastic impurity scattering being dominant at lowest $T$. 
On a closer look, one sees that the AMFL model results for  $T_\textrm{c}=0\,$K (only Fermi liquid like scattering) approaches Sommerfeld value quadratically in $T$ while the AMFL result for $T_\textrm{c}=40\,$K (with considerable marginal Fermi liquid component in the scattering) approaches Sommerfeld value more linearly in $T$. $\mathcal{L}$ therefore holds also information of the $\omega$ dependence of the scattering rate. See also Ref.~\onlinecite{tulipman22x} for further discussion of the $T$-dependence of $\mathcal{L}$ at low  $T$. 

We show in Fig.~\ref{fig_ybco_L} also the measured data for YBCO with doping p=0.12 and p=0.17. These are taken from Ref.~\onlinecite{takenaka97} and differ from the complementary $\mathcal{L}_{xy}$ measured data for YBCO with p=0.19 taken from Ref.~\onlinecite{zhang00}. 
All measured data deviate from the standard theoretical expectations, which calls for further investigation.

We note that values of $\mathcal L$ below the Sommerfeld value
can arise also due to phonon scattering \cite{lavasani2019}
or increased importance of normal vs. umklapp scattering
\cite{zhang00}.


\begin{thebibliography}{79}%
	\makeatletter
	\providecommand \@ifxundefined [1]{%
		\@ifx{#1\undefined}
	}%
	\providecommand \@ifnum [1]{%
		\ifnum #1\expandafter \@firstoftwo
		\else \expandafter \@secondoftwo
		\fi
	}%
	\providecommand \@ifx [1]{%
		\ifx #1\expandafter \@firstoftwo
		\else \expandafter \@secondoftwo
		\fi
	}%
	\providecommand \natexlab [1]{#1}%
	\providecommand \enquote  [1]{``#1''}%
	\providecommand \bibnamefont  [1]{#1}%
	\providecommand \bibfnamefont [1]{#1}%
	\providecommand \citenamefont [1]{#1}%
	\providecommand \href@noop [0]{\@secondoftwo}%
	\providecommand \href [0]{\begingroup \@sanitize@url \@href}%
	\providecommand \@href[1]{\@@startlink{#1}\@@href}%
	\providecommand \@@href[1]{\endgroup#1\@@endlink}%
	\providecommand \@sanitize@url [0]{\catcode `\\12\catcode `\$12\catcode
		`\&12\catcode `\#12\catcode `\^12\catcode `\_12\catcode `\%12\relax}%
	\providecommand \@@startlink[1]{}%
	\providecommand \@@endlink[0]{}%
	\providecommand \url  [0]{\begingroup\@sanitize@url \@url }%
	\providecommand \@url [1]{\endgroup\@href {#1}{\urlprefix }}%
	\providecommand \urlprefix  [0]{URL }%
	\providecommand \Eprint [0]{\href }%
	\providecommand \doibase [0]{http://dx.doi.org/}%
	\providecommand \selectlanguage [0]{\@gobble}%
	\providecommand \bibinfo  [0]{\@secondoftwo}%
	\providecommand \bibfield  [0]{\@secondoftwo}%
	\providecommand \translation [1]{[#1]}%
	\providecommand \BibitemOpen [0]{}%
	\providecommand \bibitemStop [0]{}%
	\providecommand \bibitemNoStop [0]{.\EOS\space}%
	\providecommand \EOS [0]{\spacefactor3000\relax}%
	\providecommand \BibitemShut  [1]{\csname bibitem#1\endcsname}%
	\let\auto@bib@innerbib\@empty
	\bibitem [{\citenamefont {Hill}\ \emph {et~al.}(2001)\citenamefont {Hill},
		\citenamefont {Proust}, \citenamefont {Taillefer}, \citenamefont {Fournier},\
		and\ \citenamefont {Greene}}]{hill01}%
	\BibitemOpen
	\bibfield  {author} {\bibinfo {author} {\bibfnamefont {R.~W.}\ \bibnamefont
			{Hill}}, \bibinfo {author} {\bibfnamefont {C.}~\bibnamefont {Proust}},
		\bibinfo {author} {\bibfnamefont {L.}~\bibnamefont {Taillefer}}, \bibinfo
		{author} {\bibfnamefont {P.}~\bibnamefont {Fournier}}, \ and\ \bibinfo
		{author} {\bibfnamefont {R.~L.}\ \bibnamefont {Greene}},\ }\href {\doibase
		10.1038/414711a} {\bibfield  {journal} {\bibinfo  {journal} {Nature}\
		}\textbf {\bibinfo {volume} {414}},\ \bibinfo {pages} {711} (\bibinfo {year}
		{2001})}\BibitemShut {NoStop}%
	\bibitem [{\citenamefont {Yamashita}\ \emph {et~al.}(2010)\citenamefont
		{Yamashita}, \citenamefont {Nakata}, \citenamefont {Senshu}, \citenamefont
		{Nagata}, \citenamefont {Yamamoto}, \citenamefont {Kato}, \citenamefont
		{Shibauchi},\ and\ \citenamefont {Matsuda}}]{yamashita10}%
	\BibitemOpen
	\bibfield  {author} {\bibinfo {author} {\bibfnamefont {M.}~\bibnamefont
			{Yamashita}}, \bibinfo {author} {\bibfnamefont {N.}~\bibnamefont {Nakata}},
		\bibinfo {author} {\bibfnamefont {Y.}~\bibnamefont {Senshu}}, \bibinfo
		{author} {\bibfnamefont {M.}~\bibnamefont {Nagata}}, \bibinfo {author}
		{\bibfnamefont {H.~M.}\ \bibnamefont {Yamamoto}}, \bibinfo {author}
		{\bibfnamefont {R.}~\bibnamefont {Kato}}, \bibinfo {author} {\bibfnamefont
			{T.}~\bibnamefont {Shibauchi}}, \ and\ \bibinfo {author} {\bibfnamefont
			{Y.}~\bibnamefont {Matsuda}},\ }\href {\doibase 10.1126/science.1188200}
	{\bibfield  {journal} {\bibinfo  {journal} {Science}\ }\textbf {\bibinfo
			{volume} {328}},\ \bibinfo {pages} {1246} (\bibinfo {year}
		{2010})}\BibitemShut {NoStop}%
	\bibitem [{\citenamefont {Lee}\ \emph {et~al.}(2017)\citenamefont {Lee},
		\citenamefont {Hippalgaonkar}, \citenamefont {Yang}, \citenamefont {Hong},
		\citenamefont {Ko}, \citenamefont {Suh}, \citenamefont {Liu}, \citenamefont
		{Wang}, \citenamefont {Urban}, \citenamefont {Zhang}, \citenamefont {Dames},
		\citenamefont {Hartnoll}, \citenamefont {Delaire},\ and\ \citenamefont
		{Wu}}]{lee17}%
	\BibitemOpen
	\bibfield  {author} {\bibinfo {author} {\bibfnamefont {S.}~\bibnamefont
			{Lee}}, \bibinfo {author} {\bibfnamefont {K.}~\bibnamefont {Hippalgaonkar}},
		\bibinfo {author} {\bibfnamefont {F.}~\bibnamefont {Yang}}, \bibinfo {author}
		{\bibfnamefont {J.}~\bibnamefont {Hong}}, \bibinfo {author} {\bibfnamefont
			{C.}~\bibnamefont {Ko}}, \bibinfo {author} {\bibfnamefont {J.}~\bibnamefont
			{Suh}}, \bibinfo {author} {\bibfnamefont {K.}~\bibnamefont {Liu}}, \bibinfo
		{author} {\bibfnamefont {K.}~\bibnamefont {Wang}}, \bibinfo {author}
		{\bibfnamefont {J.~J.}\ \bibnamefont {Urban}}, \bibinfo {author}
		{\bibfnamefont {X.}~\bibnamefont {Zhang}}, \bibinfo {author} {\bibfnamefont
			{C.}~\bibnamefont {Dames}}, \bibinfo {author} {\bibfnamefont {S.~A.}\
			\bibnamefont {Hartnoll}}, \bibinfo {author} {\bibfnamefont {O.}~\bibnamefont
			{Delaire}}, \ and\ \bibinfo {author} {\bibfnamefont {J.}~\bibnamefont {Wu}},\
	}\href {\doibase 10.1126/science.aag0410} {\bibfield  {journal} {\bibinfo
			{journal} {Science}\ }\textbf {\bibinfo {volume} {355}},\ \bibinfo {pages}
		{371} (\bibinfo {year} {2017})}\BibitemShut {NoStop}%
	\bibitem [{\citenamefont {Brown}\ \emph {et~al.}(2019)\citenamefont {Brown},
		\citenamefont {Mitra}, \citenamefont {Guardado-Sanchez}, \citenamefont
		{Nourafkan}, \citenamefont {Reymbaut}, \citenamefont {H{\'e}bert},
		\citenamefont {Bergeron}, \citenamefont {Tremblay}, \citenamefont {Kokalj},
		\citenamefont {Huse}, \citenamefont {Schau{\ss}},\ and\ \citenamefont
		{Bakr}}]{brown19}%
	\BibitemOpen
	\bibfield  {author} {\bibinfo {author} {\bibfnamefont {P.~T.}\ \bibnamefont
			{Brown}}, \bibinfo {author} {\bibfnamefont {D.}~\bibnamefont {Mitra}},
		\bibinfo {author} {\bibfnamefont {E.}~\bibnamefont {Guardado-Sanchez}},
		\bibinfo {author} {\bibfnamefont {R.}~\bibnamefont {Nourafkan}}, \bibinfo
		{author} {\bibfnamefont {A.}~\bibnamefont {Reymbaut}}, \bibinfo {author}
		{\bibfnamefont {C.-D.}\ \bibnamefont {H{\'e}bert}}, \bibinfo {author}
		{\bibfnamefont {S.}~\bibnamefont {Bergeron}}, \bibinfo {author}
		{\bibfnamefont {A.-M.~S.}\ \bibnamefont {Tremblay}}, \bibinfo {author}
		{\bibfnamefont {J.}~\bibnamefont {Kokalj}}, \bibinfo {author} {\bibfnamefont
			{D.~A.}\ \bibnamefont {Huse}}, \bibinfo {author} {\bibfnamefont
			{P.}~\bibnamefont {Schau{\ss}}}, \ and\ \bibinfo {author} {\bibfnamefont
			{W.~S.}\ \bibnamefont {Bakr}},\ }\href {\doibase 10.1126/science.aat4134}
	{\bibfield  {journal} {\bibinfo  {journal} {Science}\ }\textbf {\bibinfo
			{volume} {363}},\ \bibinfo {pages} {379} (\bibinfo {year}
		{2019})}\BibitemShut {NoStop}%
	\bibitem [{\citenamefont {Kokalj}(2017)}]{kokalj17}%
	\BibitemOpen
	\bibfield  {author} {\bibinfo {author} {\bibfnamefont {J.}~\bibnamefont
			{Kokalj}},\ }\href {\doibase 10.1103/PhysRevB.95.041110} {\bibfield
		{journal} {\bibinfo  {journal} {Phys. Rev. B}\ }\textbf {\bibinfo {volume}
			{95}},\ \bibinfo {pages} {041110} (\bibinfo {year} {2017})}\BibitemShut
	{NoStop}%
	\bibitem [{\citenamefont {Perepelitsky}\ \emph {et~al.}(2016)\citenamefont
		{Perepelitsky}, \citenamefont {Galatas}, \citenamefont {Mravlje},
		\citenamefont {\ifmmode~\check{Z}\else \v{Z}\fi{}itko}, \citenamefont
		{Khatami}, \citenamefont {Shastry},\ and\ \citenamefont
		{Georges}}]{perepelitsky16}%
	\BibitemOpen
	\bibfield  {author} {\bibinfo {author} {\bibfnamefont {E.}~\bibnamefont
			{Perepelitsky}}, \bibinfo {author} {\bibfnamefont {A.}~\bibnamefont
			{Galatas}}, \bibinfo {author} {\bibfnamefont {J.}~\bibnamefont {Mravlje}},
		\bibinfo {author} {\bibfnamefont {R.}~\bibnamefont {\ifmmode~\check{Z}\else
				\v{Z}\fi{}itko}}, \bibinfo {author} {\bibfnamefont {E.}~\bibnamefont
			{Khatami}}, \bibinfo {author} {\bibfnamefont {B.~S.}\ \bibnamefont
			{Shastry}}, \ and\ \bibinfo {author} {\bibfnamefont {A.}~\bibnamefont
			{Georges}},\ }\href {\doibase 10.1103/PhysRevB.94.235115} {\bibfield
		{journal} {\bibinfo  {journal} {Phys. Rev. B}\ }\textbf {\bibinfo {volume}
			{94}},\ \bibinfo {pages} {235115} (\bibinfo {year} {2016})}\BibitemShut
	{NoStop}%
	\bibitem [{\citenamefont {Vu\ifmmode \check{c}\else \v{c}\fi{}i\ifmmode
			\check{c}\else \v{c}\fi{}evi\ifmmode~\acute{c}\else \'{c}\fi{}}\ \emph
		{et~al.}(2019)\citenamefont {Vu\ifmmode \check{c}\else \v{c}\fi{}i\ifmmode
			\check{c}\else \v{c}\fi{}evi\ifmmode~\acute{c}\else \'{c}\fi{}},
		\citenamefont {Kokalj}, \citenamefont {\ifmmode~\check{Z}\else
			\v{Z}\fi{}itko}, \citenamefont {Wentzell}, \citenamefont
		{Tanaskovi\ifmmode~\acute{c}\else \'{c}\fi{}},\ and\ \citenamefont
		{Mravlje}}]{vucicevic19}%
	\BibitemOpen
	\bibfield  {author} {\bibinfo {author} {\bibfnamefont {J.}~\bibnamefont
			{Vu\ifmmode \check{c}\else \v{c}\fi{}i\ifmmode \check{c}\else
				\v{c}\fi{}evi\ifmmode~\acute{c}\else \'{c}\fi{}}}, \bibinfo {author}
		{\bibfnamefont {J.}~\bibnamefont {Kokalj}}, \bibinfo {author} {\bibfnamefont
			{R.}~\bibnamefont {\ifmmode~\check{Z}\else \v{Z}\fi{}itko}}, \bibinfo
		{author} {\bibfnamefont {N.}~\bibnamefont {Wentzell}}, \bibinfo {author}
		{\bibfnamefont {D.}~\bibnamefont {Tanaskovi\ifmmode~\acute{c}\else
				\'{c}\fi{}}}, \ and\ \bibinfo {author} {\bibfnamefont {J.}~\bibnamefont
			{Mravlje}},\ }\href {\doibase 10.1103/PhysRevLett.123.036601} {\bibfield
		{journal} {\bibinfo  {journal} {Phys. Rev. Lett.}\ }\textbf {\bibinfo
			{volume} {123}},\ \bibinfo {pages} {036601} (\bibinfo {year}
		{2019})}\BibitemShut {NoStop}%
	\bibitem [{\citenamefont {Vrani\ifmmode~\acute{c}\else \'{c}\fi{}}\ \emph
		{et~al.}(2020)\citenamefont {Vrani\ifmmode~\acute{c}\else \'{c}\fi{}},
		\citenamefont {Vu\ifmmode \check{c}\else \v{c}\fi{}i\ifmmode \check{c}\else
			\v{c}\fi{}evi\ifmmode~\acute{c}\else \'{c}\fi{}}, \citenamefont {Kokalj},
		\citenamefont {Skolimowski}, \citenamefont {\ifmmode~\check{Z}\else
			\v{Z}\fi{}itko}, \citenamefont {Mravlje},\ and\ \citenamefont
		{Tanaskovi\ifmmode~\acute{c}\else \'{c}\fi{}}}]{vranic20}%
	\BibitemOpen
	\bibfield  {author} {\bibinfo {author} {\bibfnamefont {A.}~\bibnamefont
			{Vrani\ifmmode~\acute{c}\else \'{c}\fi{}}}, \bibinfo {author} {\bibfnamefont
			{J.}~\bibnamefont {Vu\ifmmode \check{c}\else \v{c}\fi{}i\ifmmode
				\check{c}\else \v{c}\fi{}evi\ifmmode~\acute{c}\else \'{c}\fi{}}}, \bibinfo
		{author} {\bibfnamefont {J.}~\bibnamefont {Kokalj}}, \bibinfo {author}
		{\bibfnamefont {J.}~\bibnamefont {Skolimowski}}, \bibinfo {author}
		{\bibfnamefont {R.}~\bibnamefont {\ifmmode~\check{Z}\else \v{Z}\fi{}itko}},
		\bibinfo {author} {\bibfnamefont {J.}~\bibnamefont {Mravlje}}, \ and\
		\bibinfo {author} {\bibfnamefont {D.}~\bibnamefont
			{Tanaskovi\ifmmode~\acute{c}\else \'{c}\fi{}}},\ }\href {\doibase
		10.1103/PhysRevB.102.115142} {\bibfield  {journal} {\bibinfo  {journal}
			{Phys. Rev. B}\ }\textbf {\bibinfo {volume} {102}},\ \bibinfo {pages}
		{115142} (\bibinfo {year} {2020})}\BibitemShut {NoStop}%
	\bibitem [{\citenamefont {Huang}\ \emph {et~al.}(2019)\citenamefont {Huang},
		\citenamefont {Sheppard}, \citenamefont {Moritz},\ and\ \citenamefont
		{Devereaux}}]{huang19}%
	\BibitemOpen
	\bibfield  {author} {\bibinfo {author} {\bibfnamefont {E.~W.}\ \bibnamefont
			{Huang}}, \bibinfo {author} {\bibfnamefont {R.}~\bibnamefont {Sheppard}},
		\bibinfo {author} {\bibfnamefont {B.}~\bibnamefont {Moritz}}, \ and\ \bibinfo
		{author} {\bibfnamefont {T.~P.}\ \bibnamefont {Devereaux}},\ }\href {\doibase
		10.1126/science.aau7063} {\bibfield  {journal} {\bibinfo  {journal}
			{Science}\ }\textbf {\bibinfo {volume} {366}},\ \bibinfo {pages} {987}
		(\bibinfo {year} {2019})}\BibitemShut {NoStop}%
	\bibitem [{\citenamefont {Gunnarsson}\ \emph {et~al.}(2003)\citenamefont
		{Gunnarsson}, \citenamefont {Calandra},\ and\ \citenamefont
		{Han}}]{gunnarsson03}%
	\BibitemOpen
	\bibfield  {author} {\bibinfo {author} {\bibfnamefont {O.}~\bibnamefont
			{Gunnarsson}}, \bibinfo {author} {\bibfnamefont {M.}~\bibnamefont
			{Calandra}}, \ and\ \bibinfo {author} {\bibfnamefont {J.~E.}\ \bibnamefont
			{Han}},\ }\href {\doibase 10.1103/RevModPhys.75.1085} {\bibfield  {journal}
		{\bibinfo  {journal} {Rev. Mod. Phys.}\ }\textbf {\bibinfo {volume} {75}},\
		\bibinfo {pages} {1085} (\bibinfo {year} {2003})}\BibitemShut {NoStop}%
	\bibitem [{\citenamefont {Nichols}\ \emph {et~al.}(2019)\citenamefont
		{Nichols}, \citenamefont {Cheuk}, \citenamefont {Okan}, \citenamefont
		{Hartke}, \citenamefont {Mendez}, \citenamefont {Senthil}, \citenamefont
		{Khatami}, \citenamefont {Zhang},\ and\ \citenamefont
		{Zwierlein}}]{nichols19}%
	\BibitemOpen
	\bibfield  {author} {\bibinfo {author} {\bibfnamefont {M.~A.}\ \bibnamefont
			{Nichols}}, \bibinfo {author} {\bibfnamefont {L.~W.}\ \bibnamefont {Cheuk}},
		\bibinfo {author} {\bibfnamefont {M.}~\bibnamefont {Okan}}, \bibinfo {author}
		{\bibfnamefont {T.~R.}\ \bibnamefont {Hartke}}, \bibinfo {author}
		{\bibfnamefont {E.}~\bibnamefont {Mendez}}, \bibinfo {author} {\bibfnamefont
			{T.}~\bibnamefont {Senthil}}, \bibinfo {author} {\bibfnamefont
			{E.}~\bibnamefont {Khatami}}, \bibinfo {author} {\bibfnamefont
			{H.}~\bibnamefont {Zhang}}, \ and\ \bibinfo {author} {\bibfnamefont {M.~W.}\
			\bibnamefont {Zwierlein}},\ }\href {\doibase 10.1126/science.aat4387}
	{\bibfield  {journal} {\bibinfo  {journal} {Science}\ }\textbf {\bibinfo
			{volume} {363}},\ \bibinfo {pages} {383} (\bibinfo {year}
		{2019})}\BibitemShut {NoStop}%
	\bibitem [{\citenamefont {Ulaga}\ \emph {et~al.}(2021)\citenamefont {Ulaga},
		\citenamefont {Mravlje},\ and\ \citenamefont {Kokalj}}]{ulaga21}%
	\BibitemOpen
	\bibfield  {author} {\bibinfo {author} {\bibfnamefont {M.}~\bibnamefont
			{Ulaga}}, \bibinfo {author} {\bibfnamefont {J.}~\bibnamefont {Mravlje}}, \
		and\ \bibinfo {author} {\bibfnamefont {J.}~\bibnamefont {Kokalj}},\ }\href
	{\doibase 10.1103/PhysRevB.103.155123} {\bibfield  {journal} {\bibinfo
			{journal} {Phys. Rev. B}\ }\textbf {\bibinfo {volume} {103}},\ \bibinfo
		{pages} {155123} (\bibinfo {year} {2021})}\BibitemShut {NoStop}%
	\bibitem [{\citenamefont {Zhang}\ \emph {et~al.}(2017)\citenamefont {Zhang},
		\citenamefont {Levenson-Falk}, \citenamefont {Ramshaw}, \citenamefont {Bonn},
		\citenamefont {Liang}, \citenamefont {Hardy}, \citenamefont {Hartnoll},\ and\
		\citenamefont {Kapitulnik}}]{zhang17}%
	\BibitemOpen
	\bibfield  {author} {\bibinfo {author} {\bibfnamefont {J.}~\bibnamefont
			{Zhang}}, \bibinfo {author} {\bibfnamefont {E.~M.}\ \bibnamefont
			{Levenson-Falk}}, \bibinfo {author} {\bibfnamefont {B.~J.}\ \bibnamefont
			{Ramshaw}}, \bibinfo {author} {\bibfnamefont {D.~A.}\ \bibnamefont {Bonn}},
		\bibinfo {author} {\bibfnamefont {R.}~\bibnamefont {Liang}}, \bibinfo
		{author} {\bibfnamefont {W.~N.}\ \bibnamefont {Hardy}}, \bibinfo {author}
		{\bibfnamefont {S.~A.}\ \bibnamefont {Hartnoll}}, \ and\ \bibinfo {author}
		{\bibfnamefont {A.}~\bibnamefont {Kapitulnik}},\ }\href {\doibase
		10.1073/pnas.1703416114} {\bibfield  {journal} {\bibinfo  {journal} {Proc.
				Natl. Acad. Sci.}\ }\textbf {\bibinfo {volume} {114}},\ \bibinfo {pages}
		{5378} (\bibinfo {year} {2017})}\BibitemShut {NoStop}%
	\bibitem [{\citenamefont {Zhang}\ \emph {et~al.}(2019)\citenamefont {Zhang},
		\citenamefont {Kountz}, \citenamefont {Levenson-Falk}, \citenamefont {Song},
		\citenamefont {Greene},\ and\ \citenamefont {Kapitulnik}}]{zhang19}%
	\BibitemOpen
	\bibfield  {author} {\bibinfo {author} {\bibfnamefont {J.}~\bibnamefont
			{Zhang}}, \bibinfo {author} {\bibfnamefont {E.~D.}\ \bibnamefont {Kountz}},
		\bibinfo {author} {\bibfnamefont {E.~M.}\ \bibnamefont {Levenson-Falk}},
		\bibinfo {author} {\bibfnamefont {D.}~\bibnamefont {Song}}, \bibinfo {author}
		{\bibfnamefont {R.~L.}\ \bibnamefont {Greene}}, \ and\ \bibinfo {author}
		{\bibfnamefont {A.}~\bibnamefont {Kapitulnik}},\ }\href {\doibase
		10.1103/PhysRevB.100.241114} {\bibfield  {journal} {\bibinfo  {journal}
			{Phys. Rev. B}\ }\textbf {\bibinfo {volume} {100}},\ \bibinfo {pages}
		{241114} (\bibinfo {year} {2019})}\BibitemShut {NoStop}%
	\bibitem [{\citenamefont {Martelli}\ \emph {et~al.}(2018)\citenamefont
		{Martelli}, \citenamefont {Jim\'enez}, \citenamefont {Continentino},
		\citenamefont {Baggio-Saitovitch},\ and\ \citenamefont
		{Behnia}}]{martelli18}%
	\BibitemOpen
	\bibfield  {author} {\bibinfo {author} {\bibfnamefont {V.}~\bibnamefont
			{Martelli}}, \bibinfo {author} {\bibfnamefont {J.~L.}\ \bibnamefont
			{Jim\'enez}}, \bibinfo {author} {\bibfnamefont {M.}~\bibnamefont
			{Continentino}}, \bibinfo {author} {\bibfnamefont {E.}~\bibnamefont
			{Baggio-Saitovitch}}, \ and\ \bibinfo {author} {\bibfnamefont
			{K.}~\bibnamefont {Behnia}},\ }\href {\doibase
		10.1103/PhysRevLett.120.125901} {\bibfield  {journal} {\bibinfo  {journal}
			{Phys. Rev. Lett.}\ }\textbf {\bibinfo {volume} {120}},\ \bibinfo {pages}
		{125901} (\bibinfo {year} {2018})}\BibitemShut {NoStop}%
	\bibitem [{\citenamefont {Zhang}\ \emph {et~al.}(2000)\citenamefont {Zhang},
		\citenamefont {Ong}, \citenamefont {Xu}, \citenamefont {Krishana},
		\citenamefont {Gagnon},\ and\ \citenamefont {Taillefer}}]{zhang00}%
	\BibitemOpen
	\bibfield  {author} {\bibinfo {author} {\bibfnamefont {Y.}~\bibnamefont
			{Zhang}}, \bibinfo {author} {\bibfnamefont {N.~P.}\ \bibnamefont {Ong}},
		\bibinfo {author} {\bibfnamefont {Z.~A.}\ \bibnamefont {Xu}}, \bibinfo
		{author} {\bibfnamefont {K.}~\bibnamefont {Krishana}}, \bibinfo {author}
		{\bibfnamefont {R.}~\bibnamefont {Gagnon}}, \ and\ \bibinfo {author}
		{\bibfnamefont {L.}~\bibnamefont {Taillefer}},\ }\href {\doibase
		10.1103/PhysRevLett.84.2219} {\bibfield  {journal} {\bibinfo  {journal}
			{Phys. Rev. Lett.}\ }\textbf {\bibinfo {volume} {84}},\ \bibinfo {pages}
		{2219} (\bibinfo {year} {2000})}\BibitemShut {NoStop}%
	\bibitem [{\citenamefont {Kim}\ and\ \citenamefont {P\'epin}(2009)}]{kim09}%
	\BibitemOpen
	\bibfield  {author} {\bibinfo {author} {\bibfnamefont {K.-S.}\ \bibnamefont
			{Kim}}\ and\ \bibinfo {author} {\bibfnamefont {C.}~\bibnamefont {P\'epin}},\
	}\href {\doibase 10.1103/PhysRevLett.102.156404} {\bibfield  {journal}
		{\bibinfo  {journal} {Phys. Rev. Lett.}\ }\textbf {\bibinfo {volume} {102}},\
		\bibinfo {pages} {156404} (\bibinfo {year} {2009})}\BibitemShut {NoStop}%
	\bibitem [{\citenamefont {Mahajan}\ \emph {et~al.}(2013)\citenamefont
		{Mahajan}, \citenamefont {Barkeshli},\ and\ \citenamefont
		{Hartnoll}}]{mahajan13}%
	\BibitemOpen
	\bibfield  {author} {\bibinfo {author} {\bibfnamefont {R.}~\bibnamefont
			{Mahajan}}, \bibinfo {author} {\bibfnamefont {M.}~\bibnamefont {Barkeshli}},
		\ and\ \bibinfo {author} {\bibfnamefont {S.~A.}\ \bibnamefont {Hartnoll}},\
	}\href {\doibase 10.1103/PhysRevB.88.125107} {\bibfield  {journal} {\bibinfo
			{journal} {Phys. Rev. B}\ }\textbf {\bibinfo {volume} {88}},\ \bibinfo
		{pages} {125107} (\bibinfo {year} {2013})}\BibitemShut {NoStop}%
	\bibitem [{\citenamefont {Lavasani}\ \emph {et~al.}(2019)\citenamefont
		{Lavasani}, \citenamefont {Bulmash},\ and\ \citenamefont
		{Sarma}}]{lavasani2019}%
	\BibitemOpen
	\bibfield  {author} {\bibinfo {author} {\bibfnamefont {A.}~\bibnamefont
			{Lavasani}}, \bibinfo {author} {\bibfnamefont {D.}~\bibnamefont {Bulmash}}, \
		and\ \bibinfo {author} {\bibfnamefont {S.~D.}\ \bibnamefont {Sarma}},\ }\href
	{\doibase 10.1103/PhysRevB.99.085104} {\bibfield  {journal} {\bibinfo
			{journal} {Phys. Rev. B}\ }\textbf {\bibinfo {volume} {99}},\ \bibinfo
		{pages} {085104} (\bibinfo {year} {2019})}\BibitemShut {NoStop}%
	\bibitem [{\citenamefont {Yu}\ \emph {et~al.}(1992)\citenamefont {Yu},
		\citenamefont {Salamon}, \citenamefont {Lu},\ and\ \citenamefont
		{Lee}}]{yu92}%
	\BibitemOpen
	\bibfield  {author} {\bibinfo {author} {\bibfnamefont {R.~C.}\ \bibnamefont
			{Yu}}, \bibinfo {author} {\bibfnamefont {M.~B.}\ \bibnamefont {Salamon}},
		\bibinfo {author} {\bibfnamefont {J.~P.}\ \bibnamefont {Lu}}, \ and\ \bibinfo
		{author} {\bibfnamefont {W.~C.}\ \bibnamefont {Lee}},\ }\href {\doibase
		10.1103/PhysRevLett.69.1431} {\bibfield  {journal} {\bibinfo  {journal}
			{Phys. Rev. Lett.}\ }\textbf {\bibinfo {volume} {69}},\ \bibinfo {pages}
		{1431} (\bibinfo {year} {1992})}\BibitemShut {NoStop}%
	\bibitem [{\citenamefont {Allen}\ \emph {et~al.}(1994)\citenamefont {Allen},
		\citenamefont {Du}, \citenamefont {Mihaly},\ and\ \citenamefont
		{Forro}}]{allen94}%
	\BibitemOpen
	\bibfield  {author} {\bibinfo {author} {\bibfnamefont {P.~B.}\ \bibnamefont
			{Allen}}, \bibinfo {author} {\bibfnamefont {X.}~\bibnamefont {Du}}, \bibinfo
		{author} {\bibfnamefont {L.}~\bibnamefont {Mihaly}}, \ and\ \bibinfo {author}
		{\bibfnamefont {L.}~\bibnamefont {Forro}},\ }\href {\doibase
		10.1103/PhysRevB.49.9073} {\bibfield  {journal} {\bibinfo  {journal} {Phys.
				Rev. B}\ }\textbf {\bibinfo {volume} {49}},\ \bibinfo {pages} {9073}
		(\bibinfo {year} {1994})}\BibitemShut {NoStop}%
	\bibitem [{\citenamefont {Yan}\ \emph {et~al.}(2004)\citenamefont {Yan},
		\citenamefont {Zhou},\ and\ \citenamefont {Goodenough}}]{yan04}%
	\BibitemOpen
	\bibfield  {author} {\bibinfo {author} {\bibfnamefont {J.-Q.}\ \bibnamefont
			{Yan}}, \bibinfo {author} {\bibfnamefont {J.-S.}\ \bibnamefont {Zhou}}, \
		and\ \bibinfo {author} {\bibfnamefont {J.~B.}\ \bibnamefont {Goodenough}},\
	}\href {\doibase 10.1088/1367-2630/6/1/143} {\bibfield  {journal} {\bibinfo
			{journal} {New J. Sci.}\ }\textbf {\bibinfo {volume} {6}},\ \bibinfo {pages}
		{143} (\bibinfo {year} {2004})}\BibitemShut {NoStop}%
	\bibitem [{\citenamefont {Hess}\ \emph {et~al.}(2003)\citenamefont {Hess},
		\citenamefont {B\"uchner}, \citenamefont {Ammerahl}, \citenamefont
		{Colonescu}, \citenamefont {Heidrich-Meisner}, \citenamefont {Brenig},\ and\
		\citenamefont {Revcolevschi}}]{hess03}%
	\BibitemOpen
	\bibfield  {author} {\bibinfo {author} {\bibfnamefont {C.}~\bibnamefont
			{Hess}}, \bibinfo {author} {\bibfnamefont {B.}~\bibnamefont {B\"uchner}},
		\bibinfo {author} {\bibfnamefont {U.}~\bibnamefont {Ammerahl}}, \bibinfo
		{author} {\bibfnamefont {L.}~\bibnamefont {Colonescu}}, \bibinfo {author}
		{\bibfnamefont {F.}~\bibnamefont {Heidrich-Meisner}}, \bibinfo {author}
		{\bibfnamefont {W.}~\bibnamefont {Brenig}}, \ and\ \bibinfo {author}
		{\bibfnamefont {A.}~\bibnamefont {Revcolevschi}},\ }\href {\doibase
		10.1103/PhysRevLett.90.197002} {\bibfield  {journal} {\bibinfo  {journal}
			{Phys. Rev. Lett.}\ }\textbf {\bibinfo {volume} {90}},\ \bibinfo {pages}
		{197002} (\bibinfo {year} {2003})}\BibitemShut {NoStop}%
	\bibitem [{\citenamefont {Mousatov}\ and\ \citenamefont
		{Hartnoll}(2020)}]{mousatov20}%
	\BibitemOpen
	\bibfield  {author} {\bibinfo {author} {\bibfnamefont {C.~H.}\ \bibnamefont
			{Mousatov}}\ and\ \bibinfo {author} {\bibfnamefont {S.~A.}\ \bibnamefont
			{Hartnoll}},\ }\href {\doibase 10.1038/s41567-020-0828-6} {\bibfield
		{journal} {\bibinfo  {journal} {Nat. Phys.}\ }\textbf {\bibinfo {volume}
			{16}},\ \bibinfo {pages} {579} (\bibinfo {year} {2020})}\BibitemShut
	{NoStop}%
	\bibitem [{\citenamefont {Zotos}(1999)}]{zotos98}%
	\BibitemOpen
	\bibfield  {author} {\bibinfo {author} {\bibfnamefont {X.}~\bibnamefont
			{Zotos}},\ }\href {\doibase 10.1103/PhysRevLett.82.1764} {\bibfield
		{journal} {\bibinfo  {journal} {Phys. Rev. Lett.}\ }\textbf {\bibinfo
			{volume} {82}},\ \bibinfo {pages} {1764} (\bibinfo {year}
		{1999})}\BibitemShut {NoStop}%
	\bibitem [{\citenamefont {Zotos}(2005)}]{zotos05}%
	\BibitemOpen
	\bibfield  {author} {\bibinfo {author} {\bibfnamefont {X.}~\bibnamefont
			{Zotos}},\ }\href {\doibase 10.1143/JPSJS.74S.173} {\bibfield  {journal}
		{\bibinfo  {journal} {J. Phys. Soc. Japan}\ }\textbf {\bibinfo {volume}
			{74}},\ \bibinfo {pages} {173} (\bibinfo {year} {2005})}\BibitemShut
	{NoStop}%
	\bibitem [{\citenamefont {Hess}(2007)}]{hess07}%
	\BibitemOpen
	\bibfield  {author} {\bibinfo {author} {\bibfnamefont {C.}~\bibnamefont
			{Hess}},\ }\href {\doibase 10.1140/epjst/e2007-00363-8} {\bibfield  {journal}
		{\bibinfo  {journal} {Eur. Phys. J. Spec. Top.}\ }\textbf {\bibinfo {volume}
			{151}},\ \bibinfo {pages} {73} (\bibinfo {year} {2007})}\BibitemShut
	{NoStop}%
	\bibitem [{\citenamefont {Karrasch}(2017)}]{karrasch17}%
	\BibitemOpen
	\bibfield  {author} {\bibinfo {author} {\bibfnamefont {C.}~\bibnamefont
			{Karrasch}},\ }\href {\doibase 10.1088/1367-2630/aa631a} {\bibfield
		{journal} {\bibinfo  {journal} {New J. Phys.}\ }\textbf {\bibinfo {volume}
			{19}},\ \bibinfo {pages} {033027} (\bibinfo {year} {2017})}\BibitemShut
	{NoStop}%
	\bibitem [{\citenamefont {Wang}\ \emph
		{et~al.}(2022{\natexlab{a}})\citenamefont {Wang}, \citenamefont {Ding},
		\citenamefont {Moritz}, \citenamefont {Huang},\ and\ \citenamefont
		{Devereaux}}]{wang21}%
	\BibitemOpen
	\bibfield  {author} {\bibinfo {author} {\bibfnamefont {W.~O.}\ \bibnamefont
			{Wang}}, \bibinfo {author} {\bibfnamefont {J.~K.}\ \bibnamefont {Ding}},
		\bibinfo {author} {\bibfnamefont {B.}~\bibnamefont {Moritz}}, \bibinfo
		{author} {\bibfnamefont {E.~W.}\ \bibnamefont {Huang}}, \ and\ \bibinfo
		{author} {\bibfnamefont {T.~P.}\ \bibnamefont {Devereaux}},\ }\href {\doibase
		10.1103/PhysRevB.105.L161103} {\bibfield  {journal} {\bibinfo  {journal}
			{Phys. Rev. B}\ }\textbf {\bibinfo {volume} {105}},\ \bibinfo {pages}
		{L161103} (\bibinfo {year} {2022}{\natexlab{a}})}\BibitemShut {NoStop}%
	\bibitem [{\citenamefont {Kiely}\ and\ \citenamefont
		{Mueller}(2021)}]{kiely21}%
	\BibitemOpen
	\bibfield  {author} {\bibinfo {author} {\bibfnamefont {T.~G.}\ \bibnamefont
			{Kiely}}\ and\ \bibinfo {author} {\bibfnamefont {E.~J.}\ \bibnamefont
			{Mueller}},\ }\href {\doibase 10.1103/PhysRevB.104.165143} {\bibfield
		{journal} {\bibinfo  {journal} {Phys. Rev. B}\ }\textbf {\bibinfo {volume}
			{104}},\ \bibinfo {pages} {165143} (\bibinfo {year} {2021})}\BibitemShut
	{NoStop}%
	\bibitem [{\citenamefont {Schneider}\ \emph {et~al.}(2012)\citenamefont
		{Schneider}, \citenamefont {Hackerm{\"u}ller}, \citenamefont {Ronzheimer},
		\citenamefont {Will}, \citenamefont {Braun}, \citenamefont {Best},
		\citenamefont {Bloch}, \citenamefont {Demler}, \citenamefont {Mandt},
		\citenamefont {Rasch},\ and\ \citenamefont {Rosch}}]{schneider12}%
	\BibitemOpen
	\bibfield  {author} {\bibinfo {author} {\bibfnamefont {U.}~\bibnamefont
			{Schneider}}, \bibinfo {author} {\bibfnamefont {L.}~\bibnamefont
			{Hackerm{\"u}ller}}, \bibinfo {author} {\bibfnamefont {J.~P.}\ \bibnamefont
			{Ronzheimer}}, \bibinfo {author} {\bibfnamefont {S.}~\bibnamefont {Will}},
		\bibinfo {author} {\bibfnamefont {S.}~\bibnamefont {Braun}}, \bibinfo
		{author} {\bibfnamefont {T.}~\bibnamefont {Best}}, \bibinfo {author}
		{\bibfnamefont {I.}~\bibnamefont {Bloch}}, \bibinfo {author} {\bibfnamefont
			{E.}~\bibnamefont {Demler}}, \bibinfo {author} {\bibfnamefont
			{S.}~\bibnamefont {Mandt}}, \bibinfo {author} {\bibfnamefont
			{D.}~\bibnamefont {Rasch}}, \ and\ \bibinfo {author} {\bibfnamefont
			{A.}~\bibnamefont {Rosch}},\ }\href {\doibase 10.1038/nphys2205} {\bibfield
		{journal} {\bibinfo  {journal} {Nat. Phys.}\ }\textbf {\bibinfo {volume}
			{8}},\ \bibinfo {pages} {213} (\bibinfo {year} {2012})}\BibitemShut {NoStop}%
	\bibitem [{\citenamefont {Guardado-Sanchez}\ \emph {et~al.}(2020)\citenamefont
		{Guardado-Sanchez}, \citenamefont {Morningstar}, \citenamefont {Spar},
		\citenamefont {Brown}, \citenamefont {Huse},\ and\ \citenamefont
		{Bakr}}]{gardadosanchez19}%
	\BibitemOpen
	\bibfield  {author} {\bibinfo {author} {\bibfnamefont {E.}~\bibnamefont
			{Guardado-Sanchez}}, \bibinfo {author} {\bibfnamefont {A.}~\bibnamefont
			{Morningstar}}, \bibinfo {author} {\bibfnamefont {B.~M.}\ \bibnamefont
			{Spar}}, \bibinfo {author} {\bibfnamefont {P.~T.}\ \bibnamefont {Brown}},
		\bibinfo {author} {\bibfnamefont {D.~A.}\ \bibnamefont {Huse}}, \ and\
		\bibinfo {author} {\bibfnamefont {W.~S.}\ \bibnamefont {Bakr}},\ }\href
	{\doibase 10.1103/PhysRevX.10.011042} {\bibfield  {journal} {\bibinfo
			{journal} {Phys. Rev. X}\ }\textbf {\bibinfo {volume} {10}},\ \bibinfo
		{pages} {011042} (\bibinfo {year} {2020})}\BibitemShut {NoStop}%
	\bibitem [{\citenamefont {Mravlje}\ \emph {et~al.}(2022)\citenamefont
		{Mravlje}, \citenamefont {Ulaga},\ and\ \citenamefont {Kokalj}}]{mravlje22}%
	\BibitemOpen
	\bibfield  {author} {\bibinfo {author} {\bibfnamefont {J.}~\bibnamefont
			{Mravlje}}, \bibinfo {author} {\bibfnamefont {M.}~\bibnamefont {Ulaga}}, \
		and\ \bibinfo {author} {\bibfnamefont {J.}~\bibnamefont {Kokalj}},\ }\href
	{\doibase 10.1103/PhysRevResearch.4.023197} {\bibfield  {journal} {\bibinfo
			{journal} {Phys. Rev. Research}\ }\textbf {\bibinfo {volume} {4}},\ \bibinfo
		{pages} {023197} (\bibinfo {year} {2022})}\BibitemShut {NoStop}%
	\bibitem [{\citenamefont {Kokalj}\ and\ \citenamefont
		{McKenzie}(2011)}]{kokalj11}%
	\BibitemOpen
	\bibfield  {author} {\bibinfo {author} {\bibfnamefont {J.}~\bibnamefont
			{Kokalj}}\ and\ \bibinfo {author} {\bibfnamefont {R.~H.}\ \bibnamefont
			{McKenzie}},\ }\href {\doibase 10.1103/PhysRevLett.107.147001} {\bibfield
		{journal} {\bibinfo  {journal} {Phys. Rev. Lett.}\ }\textbf {\bibinfo
			{volume} {107}},\ \bibinfo {pages} {147001} (\bibinfo {year}
		{2011})}\BibitemShut {NoStop}%
	\bibitem [{\citenamefont {Kokalj}\ \emph {et~al.}(2012)\citenamefont {Kokalj},
		\citenamefont {Hussey},\ and\ \citenamefont {McKenzie}}]{kokalj12}%
	\BibitemOpen
	\bibfield  {author} {\bibinfo {author} {\bibfnamefont {J.}~\bibnamefont
			{Kokalj}}, \bibinfo {author} {\bibfnamefont {N.~E.}\ \bibnamefont {Hussey}},
		\ and\ \bibinfo {author} {\bibfnamefont {R.~H.}\ \bibnamefont {McKenzie}},\
	}\href {\doibase 10.1103/PhysRevB.86.045132} {\bibfield  {journal} {\bibinfo
			{journal} {Phys. Rev. B}\ }\textbf {\bibinfo {volume} {86}},\ \bibinfo
		{pages} {045132} (\bibinfo {year} {2012})}\BibitemShut {NoStop}%
	\bibitem [{\citenamefont {Jakli\v{c}}\ and\ \citenamefont
		{Prelov\v{s}ek}(2000)}]{jaklic00}%
	\BibitemOpen
	\bibfield  {author} {\bibinfo {author} {\bibfnamefont {J.}~\bibnamefont
			{Jakli\v{c}}}\ and\ \bibinfo {author} {\bibfnamefont {P.}~\bibnamefont
			{Prelov\v{s}ek}},\ }\href {\doibase 10.1080/000187300243381} {\bibfield
		{journal} {\bibinfo  {journal} {Adv. Phys.}\ }\textbf {\bibinfo {volume}
			{49}},\ \bibinfo {pages} {1} (\bibinfo {year} {2000})}\BibitemShut {NoStop}%
	\bibitem [{\citenamefont {{Prelov\v{s}ek}}\ and\ \citenamefont
		{{Bon\v{c}a}}(2013)}]{prelovsek13}%
	\BibitemOpen
	\bibfield  {author} {\bibinfo {author} {\bibfnamefont {P.}~\bibnamefont
			{{Prelov\v{s}ek}}}\ and\ \bibinfo {author} {\bibfnamefont {J.}~\bibnamefont
			{{Bon\v{c}a}}},\ }\href {https://books.google.si/books?id=Be4\_AAAAQBAJ}
	{\emph {\bibinfo {title} {Strongly Correlated Systems: Numerical Methods}}},\
	edited by\ \bibinfo {editor} {\bibfnamefont {A.}~\bibnamefont {Avella}}\ and\
	\bibinfo {editor} {\bibfnamefont {F.}~\bibnamefont {Mancini}},\ Springer
	Series in Solid-State Sciences\ (\bibinfo  {publisher} {Springer Berlin
		Heidelberg},\ \bibinfo {year} {2013})\BibitemShut {NoStop}%
	\bibitem [{\citenamefont {Kokalj}\ and\ \citenamefont
		{McKenzie}(2013)}]{kokalj13}%
	\BibitemOpen
	\bibfield  {author} {\bibinfo {author} {\bibfnamefont {J.}~\bibnamefont
			{Kokalj}}\ and\ \bibinfo {author} {\bibfnamefont {R.~H.}\ \bibnamefont
			{McKenzie}},\ }\href {\doibase 10.1103/PhysRevLett.110.206402} {\bibfield
		{journal} {\bibinfo  {journal} {Phys. Rev. Lett.}\ }\textbf {\bibinfo
			{volume} {110}},\ \bibinfo {pages} {206402} (\bibinfo {year}
		{2013})}\BibitemShut {NoStop}%
	\bibitem [{\citenamefont {\v{Z}itko}\ and\ \citenamefont
		{Pruschke}(2009)}]{zitko2009energy}%
	\BibitemOpen
	\bibfield  {author} {\bibinfo {author} {\bibfnamefont {R.}~\bibnamefont
			{\v{Z}itko}}\ and\ \bibinfo {author} {\bibfnamefont {T.}~\bibnamefont
			{Pruschke}},\ }\href {\doibase 10.1103/PhysRevB.79.085106} {\bibfield
		{journal} {\bibinfo  {journal} {Phys. Rev. B}\ }\textbf {\bibinfo {volume}
			{79}},\ \bibinfo {pages} {085106} (\bibinfo {year} {2009})}\BibitemShut
	{NoStop}%
	\bibitem [{\citenamefont {Zitko}(2021)}]{zitko_rok_2021_4841076}%
	\BibitemOpen
	\bibfield  {author} {\bibinfo {author} {\bibfnamefont {R.}~\bibnamefont
			{Zitko}},\ }\href {\doibase 10.5281/zenodo.4841076} {\enquote {\bibinfo
			{title} {Nrg ljubljana},}\ } (\bibinfo {year} {2021})\BibitemShut {NoStop}%
	\bibitem [{Note1()}]{Note1}%
	\BibitemOpen
	\bibinfo {note} {We use the specific heat at fixed density or doping, which
		is in contrast with the specific heat at fixed chemical potential calculated
		in Refs.~\protect \rev@citealp {jaklic00,bonca03,kokalj13}.}\BibitemShut
	{Stop}%
	\bibitem [{\citenamefont {Prelov\ifmmode~\check{s}\else \v{s}\fi{}ek}\ \emph
		{et~al.}(2015)\citenamefont {Prelov\ifmmode~\check{s}\else \v{s}\fi{}ek},
		\citenamefont {Kokalj}, \citenamefont {Lenar\ifmmode \check{c}\else
			\v{c}\fi{}i\ifmmode~\check{c}\else \v{c}\fi{}},\ and\ \citenamefont
		{McKenzie}}]{prelovsek15}%
	\BibitemOpen
	\bibfield  {author} {\bibinfo {author} {\bibfnamefont {P.}~\bibnamefont
			{Prelov\ifmmode~\check{s}\else \v{s}\fi{}ek}}, \bibinfo {author}
		{\bibfnamefont {J.}~\bibnamefont {Kokalj}}, \bibinfo {author} {\bibfnamefont
			{Z.}~\bibnamefont {Lenar\ifmmode \check{c}\else
				\v{c}\fi{}i\ifmmode~\check{c}\else \v{c}\fi{}}}, \ and\ \bibinfo {author}
		{\bibfnamefont {R.~H.}\ \bibnamefont {McKenzie}},\ }\href {\doibase
		10.1103/PhysRevB.92.235155} {\bibfield  {journal} {\bibinfo  {journal} {Phys.
				Rev. B}\ }\textbf {\bibinfo {volume} {92}},\ \bibinfo {pages} {235155}
		(\bibinfo {year} {2015})}\BibitemShut {NoStop}%
	\bibitem [{\citenamefont {Bon\ifmmode~\check{c}\else \v{c}\fi{}a}\ and\
		\citenamefont {Prelov\ifmmode~\check{s}\else \v{s}\fi{}ek}(2003)}]{bonca03}%
	\BibitemOpen
	\bibfield  {author} {\bibinfo {author} {\bibfnamefont {J.}~\bibnamefont
			{Bon\ifmmode~\check{c}\else \v{c}\fi{}a}}\ and\ \bibinfo {author}
		{\bibfnamefont {P.}~\bibnamefont {Prelov\ifmmode~\check{s}\else
				\v{s}\fi{}ek}},\ }\href {\doibase 10.1103/PhysRevB.67.085103} {\bibfield
		{journal} {\bibinfo  {journal} {Phys. Rev. B}\ }\textbf {\bibinfo {volume}
			{67}},\ \bibinfo {pages} {085103} (\bibinfo {year} {2003})}\BibitemShut
	{NoStop}%
	\bibitem [{\citenamefont {Eskes}\ \emph {et~al.}(1994)\citenamefont {Eskes},
		\citenamefont {Ole\ifmmode~\acute{s}\else \'{s}\fi{}}, \citenamefont
		{Meinders},\ and\ \citenamefont {Stephan}}]{eskes94}%
	\BibitemOpen
	\bibfield  {author} {\bibinfo {author} {\bibfnamefont {H.}~\bibnamefont
			{Eskes}}, \bibinfo {author} {\bibfnamefont {A.~M.}\ \bibnamefont
			{Ole\ifmmode~\acute{s}\else \'{s}\fi{}}}, \bibinfo {author} {\bibfnamefont
			{M.~B.~J.}\ \bibnamefont {Meinders}}, \ and\ \bibinfo {author} {\bibfnamefont
			{W.}~\bibnamefont {Stephan}},\ }\href {\doibase 10.1103/PhysRevB.50.17980}
	{\bibfield  {journal} {\bibinfo  {journal} {Phys. Rev. B}\ }\textbf {\bibinfo
			{volume} {50}},\ \bibinfo {pages} {17980} (\bibinfo {year}
		{1994})}\BibitemShut {NoStop}%
	\bibitem [{\citenamefont {Hofmann}\ \emph {et~al.}(2003)\citenamefont
		{Hofmann}, \citenamefont {{L}orenz}, \citenamefont {Berggold}, \citenamefont
		{Gr{\"u}ninger}, \citenamefont {Freimuth}, \citenamefont {Uhrig},\ and\
		\citenamefont {Br{\"u}ck}}]{hofmann03}%
	\BibitemOpen
	\bibfield  {author} {\bibinfo {author} {\bibfnamefont {M.}~\bibnamefont
			{Hofmann}}, \bibinfo {author} {\bibfnamefont {T.}~\bibnamefont {{L}orenz}},
		\bibinfo {author} {\bibfnamefont {K.}~\bibnamefont {Berggold}}, \bibinfo
		{author} {\bibfnamefont {M.}~\bibnamefont {Gr{\"u}ninger}}, \bibinfo {author}
		{\bibfnamefont {A.}~\bibnamefont {Freimuth}}, \bibinfo {author}
		{\bibfnamefont {G.}~\bibnamefont {Uhrig}}, \ and\ \bibinfo {author}
		{\bibfnamefont {E.}~\bibnamefont {Br{\"u}ck}},\ }\href {\doibase
		10.1103/PhysRevB.67.184502} {\bibfield  {journal} {\bibinfo  {journal} {Phys.
				Rev. B}\ }\textbf {\bibinfo {volume} {67}},\ \bibinfo {pages} {184502}
		(\bibinfo {year} {2003})}\BibitemShut {NoStop}%
	\bibitem [{\citenamefont {Schnack}\ \emph {et~al.}(2018)\citenamefont
		{Schnack}, \citenamefont {Schulenburg},\ and\ \citenamefont
		{Richter}}]{schnack18}%
	\BibitemOpen
	\bibfield  {author} {\bibinfo {author} {\bibfnamefont {J.}~\bibnamefont
			{Schnack}}, \bibinfo {author} {\bibfnamefont {J.}~\bibnamefont
			{Schulenburg}}, \ and\ \bibinfo {author} {\bibfnamefont {J.}~\bibnamefont
			{Richter}},\ }\href {\doibase 10.1103/PhysRevB.98.094423} {\bibfield
		{journal} {\bibinfo  {journal} {Phys. Rev. B}\ }\textbf {\bibinfo {volume}
			{98}},\ \bibinfo {pages} {094423} (\bibinfo {year} {2018})}\BibitemShut
	{NoStop}%
	\bibitem [{\citenamefont {Sengupta}\ \emph {et~al.}(2003)\citenamefont
		{Sengupta}, \citenamefont {Sandvik},\ and\ \citenamefont
		{Singh}}]{sengupta03}%
	\BibitemOpen
	\bibfield  {author} {\bibinfo {author} {\bibfnamefont {P.}~\bibnamefont
			{Sengupta}}, \bibinfo {author} {\bibfnamefont {A.~W.}\ \bibnamefont
			{Sandvik}}, \ and\ \bibinfo {author} {\bibfnamefont {R.~R.~P.}\ \bibnamefont
			{Singh}},\ }\href {\doibase 10.1103/PhysRevB.68.094423} {\bibfield  {journal}
		{\bibinfo  {journal} {Phys. Rev. B}\ }\textbf {\bibinfo {volume} {68}},\
		\bibinfo {pages} {094423} (\bibinfo {year} {2003})}\BibitemShut {NoStop}%
	\bibitem [{\citenamefont {Makivi\ifmmode~\acute{c}\else \'{c}\fi{}}\ and\
		\citenamefont {Ding}(1991)}]{makivic91}%
	\BibitemOpen
	\bibfield  {author} {\bibinfo {author} {\bibfnamefont {M.~S.}\ \bibnamefont
			{Makivi\ifmmode~\acute{c}\else \'{c}\fi{}}}\ and\ \bibinfo {author}
		{\bibfnamefont {H.-Q.}\ \bibnamefont {Ding}},\ }\href {\doibase
		10.1103/PhysRevB.43.3562} {\bibfield  {journal} {\bibinfo  {journal} {Phys.
				Rev. B}\ }\textbf {\bibinfo {volume} {43}},\ \bibinfo {pages} {3562}
		(\bibinfo {year} {1991})}\BibitemShut {NoStop}%
	\bibitem [{\citenamefont {Chakravarty}\ \emph {et~al.}(1989)\citenamefont
		{Chakravarty}, \citenamefont {Halperin},\ and\ \citenamefont
		{Nelson}}]{chakravarty89}%
	\BibitemOpen
	\bibfield  {author} {\bibinfo {author} {\bibfnamefont {S.}~\bibnamefont
			{Chakravarty}}, \bibinfo {author} {\bibfnamefont {B.~I.}\ \bibnamefont
			{Halperin}}, \ and\ \bibinfo {author} {\bibfnamefont {D.~R.}\ \bibnamefont
			{Nelson}},\ }\href {\doibase 10.1103/PhysRevB.39.2344} {\bibfield  {journal}
		{\bibinfo  {journal} {Phys. Rev. B}\ }\textbf {\bibinfo {volume} {39}},\
		\bibinfo {pages} {2344} (\bibinfo {year} {1989})}\BibitemShut {NoStop}%
	\bibitem [{\citenamefont {Kim}\ and\ \citenamefont {Troyer}(1998)}]{kim98}%
	\BibitemOpen
	\bibfield  {author} {\bibinfo {author} {\bibfnamefont {J.-K.}\ \bibnamefont
			{Kim}}\ and\ \bibinfo {author} {\bibfnamefont {M.}~\bibnamefont {Troyer}},\
	}\href {\doibase 10.1103/PhysRevLett.80.2705} {\bibfield  {journal} {\bibinfo
			{journal} {Phys. Rev. Lett.}\ }\textbf {\bibinfo {volume} {80}},\ \bibinfo
		{pages} {2705} (\bibinfo {year} {1998})}\BibitemShut {NoStop}%
	\bibitem [{notemagnon()}]{notemagnon}%
	\BibitemOpen
	\bibinfo {note} {Average velocity is calculated as the expectation value $v =
		\int
		\left|\nabla_{\mathbf{k}}\epsilon_{\mathbf{k}}\right|n_{\mathbf{k}}\mathrm{d}^2k/\int
		n_{\mathbf{k}} \mathrm{d}^2k$, where
		$\epsilon_{\mathbf{k}}=2J\sqrt{1-\gamma_{\mathbf{k}}^2}$ is the magnon
		dispersion, $\gamma_{\mathbf{k}}=\frac{1}{2}(\cos(k_x)+\cos(k_y))$, and
		$n_{\mathbf{k}}$ is the Bose function.}\BibitemShut {Stop}%
	\bibitem [{\citenamefont {Igarashi}\ and\ \citenamefont
		{Nagao}(2005)}]{igarashi05}%
	\BibitemOpen
	\bibfield  {author} {\bibinfo {author} {\bibfnamefont {J.-i.}\ \bibnamefont
			{Igarashi}}\ and\ \bibinfo {author} {\bibfnamefont {T.}~\bibnamefont
			{Nagao}},\ }\href {\doibase 10.1103/PhysRevB.72.014403} {\bibfield  {journal}
		{\bibinfo  {journal} {Phys. Rev. B}\ }\textbf {\bibinfo {volume} {72}},\
		\bibinfo {pages} {014403} (\bibinfo {year} {2005})}\BibitemShut {NoStop}%
	\bibitem [{\citenamefont {Bon\ifmmode~\check{c}\else \v{c}\fi{}a}\ and\
		\citenamefont {Jakli\ifmmode~\check{c}\else \v{c}\fi{}}(1995)}]{bonca95}%
	\BibitemOpen
	\bibfield  {author} {\bibinfo {author} {\bibfnamefont {J.}~\bibnamefont
			{Bon\ifmmode~\check{c}\else \v{c}\fi{}a}}\ and\ \bibinfo {author}
		{\bibfnamefont {J.}~\bibnamefont {Jakli\ifmmode~\check{c}\else \v{c}\fi{}}},\
	}\href {\doibase 10.1103/PhysRevB.51.16083} {\bibfield  {journal} {\bibinfo
			{journal} {Phys. Rev. B}\ }\textbf {\bibinfo {volume} {51}},\ \bibinfo
		{pages} {16083} (\bibinfo {year} {1995})}\BibitemShut {NoStop}%
	\bibitem [{notemir()}]{notemir}%
	\BibitemOpen
	\bibinfo {note} {Obtained with similar estimate as the results for a MIR
		limit for electrical conductivity in Appendix in
		Ref.~\onlinecite{gunnarsson03}. The same estimate can be obtained also by
		using the Wiedemann-Franz law and the MIR limit for the charge
		conductivity.}\BibitemShut {Stop}%
	\bibitem [{\citenamefont {Deng}\ \emph {et~al.}(2013)\citenamefont {Deng},
		\citenamefont {Mravlje}, \citenamefont {\ifmmode~\check{Z}\else
			\v{Z}\fi{}itko}, \citenamefont {Ferrero}, \citenamefont {Kotliar},\ and\
		\citenamefont {Georges}}]{deng13}%
	\BibitemOpen
	\bibfield  {author} {\bibinfo {author} {\bibfnamefont {X.}~\bibnamefont
			{Deng}}, \bibinfo {author} {\bibfnamefont {J.}~\bibnamefont {Mravlje}},
		\bibinfo {author} {\bibfnamefont {R.}~\bibnamefont {\ifmmode~\check{Z}\else
				\v{Z}\fi{}itko}}, \bibinfo {author} {\bibfnamefont {M.}~\bibnamefont
			{Ferrero}}, \bibinfo {author} {\bibfnamefont {G.}~\bibnamefont {Kotliar}}, \
		and\ \bibinfo {author} {\bibfnamefont {A.}~\bibnamefont {Georges}},\ }\href
	{\doibase 10.1103/PhysRevLett.110.086401} {\bibfield  {journal} {\bibinfo
			{journal} {Phys. Rev. Lett.}\ }\textbf {\bibinfo {volume} {110}},\ \bibinfo
		{pages} {086401} (\bibinfo {year} {2013})}\BibitemShut {NoStop}%
	\bibitem [{\citenamefont {Krien}\ \emph {et~al.}(2019)\citenamefont {Krien},
		\citenamefont {van Loon}, \citenamefont {Katsnelson}, \citenamefont
		{Lichtenstein},\ and\ \citenamefont {Capone}}]{krien19}%
	\BibitemOpen
	\bibfield  {author} {\bibinfo {author} {\bibfnamefont {F.}~\bibnamefont
			{Krien}}, \bibinfo {author} {\bibfnamefont {E.~G. C.~P.}\ \bibnamefont {van
				Loon}}, \bibinfo {author} {\bibfnamefont {M.~I.}\ \bibnamefont {Katsnelson}},
		\bibinfo {author} {\bibfnamefont {A.~I.}\ \bibnamefont {Lichtenstein}}, \
		and\ \bibinfo {author} {\bibfnamefont {M.}~\bibnamefont {Capone}},\ }\href
	{\doibase 10.1103/PhysRevB.99.245128} {\bibfield  {journal} {\bibinfo
			{journal} {Phys. Rev. B}\ }\textbf {\bibinfo {volume} {99}},\ \bibinfo
		{pages} {245128} (\bibinfo {year} {2019})}\BibitemShut {NoStop}%
	\bibitem [{\citenamefont {Pakhira}\ and\ \citenamefont
		{McKenzie}(2015)}]{pakhira15}%
	\BibitemOpen
	\bibfield  {author} {\bibinfo {author} {\bibfnamefont {N.}~\bibnamefont
			{Pakhira}}\ and\ \bibinfo {author} {\bibfnamefont {R.~H.}\ \bibnamefont
			{McKenzie}},\ }\href {\doibase 10.1103/PhysRevB.91.075124} {\bibfield
		{journal} {\bibinfo  {journal} {Phys. Rev. B}\ }\textbf {\bibinfo {volume}
			{91}},\ \bibinfo {pages} {075124} (\bibinfo {year} {2015})}\BibitemShut
	{NoStop}%
	\bibitem [{Note2()}]{Note2}%
	\BibitemOpen
	\bibinfo {note} {Since on a square lattice the Fermi velocity is not constant
		over the Fermi surface the $v_\protect \textrm {0,F}^2$ is replaced with its
		averaged over the Fermi surface value $v_\protect \textrm {0,F}^2=\DOTSI
		\intop \ilimits@ d^2k v_{0,k}^2 \delta (\epsilon _k-\epsilon _\protect
		\textrm {F})/\DOTSI \intop \ilimits@ d^2k \delta (\epsilon _k-\epsilon
		_\protect \textrm {F})$}\BibitemShut {NoStop}%
	\bibitem [{Note3()}]{Note3}%
	\BibitemOpen
	\bibinfo {note} {Since $(-\protect \frac {\partial n_\protect \textrm
			{F}}{\partial \omega })\omega ^2$ mostly weights positive and negative
		frequencies $\omega \sim \pm 3T$ and in DMFT the $\Sigma ''(\omega )$ is not
		particle-hole symmetric (even in $\omega $), the value of $\Sigma ''(\omega
		=3T)$ should be understood as $\Sigma ''(\omega =3T)=[\Sigma ''(3T)+\Sigma
		''(-3T)]/2$}\BibitemShut {NoStop}%
	\bibitem [{\citenamefont {Pourovskii}\ \emph {et~al.}(2017)\citenamefont
		{Pourovskii}, \citenamefont {Mravlje}, \citenamefont {Georges}, \citenamefont
		{Simak},\ and\ \citenamefont {Abrikosov}}]{pourovskii17}%
	\BibitemOpen
	\bibfield  {author} {\bibinfo {author} {\bibfnamefont {L.~V.}\ \bibnamefont
			{Pourovskii}}, \bibinfo {author} {\bibfnamefont {J.}~\bibnamefont {Mravlje}},
		\bibinfo {author} {\bibfnamefont {A.}~\bibnamefont {Georges}}, \bibinfo
		{author} {\bibfnamefont {S.~I.}\ \bibnamefont {Simak}}, \ and\ \bibinfo
		{author} {\bibfnamefont {I.~A.}\ \bibnamefont {Abrikosov}},\ }\href {\doibase
		10.1088/1367-2630/aa76c9} {\bibfield  {journal} {\bibinfo  {journal} {New J.
				Phys.}\ }\textbf {\bibinfo {volume} {19}},\ \bibinfo {pages} {073022}
		(\bibinfo {year} {2017})}\BibitemShut {NoStop}%
	\bibitem [{Note4()}]{Note4}%
	\BibitemOpen
	\bibinfo {note} {The same dependencies are obtained by using constant
		diffusion constants and dependencies $\chi _\protect \textrm {c}\propto 1/T$
		and $c_\protect \textrm {el}\propto 1/T^2$.}\BibitemShut {Stop}%
	\bibitem [{\citenamefont {Wang}\ \emph
		{et~al.}(2022{\natexlab{b}})\citenamefont {Wang}, \citenamefont {Ding},
		\citenamefont {Schattner}, \citenamefont {Huang}, \citenamefont {Moritz},\
		and\ \citenamefont {Devereaux}}]{wang22x}%
	\BibitemOpen
	\bibfield  {author} {\bibinfo {author} {\bibfnamefont {W.~O.}\ \bibnamefont
			{Wang}}, \bibinfo {author} {\bibfnamefont {J.~K.}\ \bibnamefont {Ding}},
		\bibinfo {author} {\bibfnamefont {Y.}~\bibnamefont {Schattner}}, \bibinfo
		{author} {\bibfnamefont {E.~W.}\ \bibnamefont {Huang}}, \bibinfo {author}
		{\bibfnamefont {B.}~\bibnamefont {Moritz}}, \ and\ \bibinfo {author}
		{\bibfnamefont {T.~P.}\ \bibnamefont {Devereaux}},\ }\href
	{https://arxiv.org/abs/2208.09144} {\bibfield  {journal} {\bibinfo  {journal}
			{arXiv:2208.09144 [cond-mat]}\ } (\bibinfo {year}
		{2022}{\natexlab{b}})}\BibitemShut {NoStop}%
	\bibitem [{Note5()}]{Note5}%
	\BibitemOpen
	\bibinfo {note} {For LCO we used parameters $J=1550\protect \,$K, lattice
		constants $a_0=3.8\protect \,$\r A\protect \,and $c_0=13.2\protect \,$\r
		A~\cite {hess03}\protect \,with two CuO planes within $c_0$.}\BibitemShut
	{Stop}%
	\bibitem [{\citenamefont {Takenaka}\ \emph {et~al.}(1997)\citenamefont
		{Takenaka}, \citenamefont {Fukuzumi}, \citenamefont {Mizuhashi},
		\citenamefont {Uchida}, \citenamefont {Asaoka},\ and\ \citenamefont
		{Takei}}]{takenaka97}%
	\BibitemOpen
	\bibfield  {author} {\bibinfo {author} {\bibfnamefont {K.}~\bibnamefont
			{Takenaka}}, \bibinfo {author} {\bibfnamefont {Y.}~\bibnamefont {Fukuzumi}},
		\bibinfo {author} {\bibfnamefont {K.}~\bibnamefont {Mizuhashi}}, \bibinfo
		{author} {\bibfnamefont {S.}~\bibnamefont {Uchida}}, \bibinfo {author}
		{\bibfnamefont {H.}~\bibnamefont {Asaoka}}, \ and\ \bibinfo {author}
		{\bibfnamefont {H.}~\bibnamefont {Takei}},\ }\href {\doibase
		10.1103/PhysRevB.56.5654} {\bibfield  {journal} {\bibinfo  {journal} {Phys.
				Rev. B}\ }\textbf {\bibinfo {volume} {56}},\ \bibinfo {pages} {5654}
		(\bibinfo {year} {1997})}\BibitemShut {NoStop}%
	\bibitem [{Note6()}]{Note6}%
	\BibitemOpen
	\bibinfo {note} {For YBCO we used unit cell parameters $a_0=3.82\protect
		\,$\r A, $b_0=3.89\protect \,$\r A\protect \, and $c_0=11.68\protect \,$\r
		A\protect \, from Ref.~\protect \rev@citealp {bondarenko17,varshney11}
		together with $t=0.3$ eV and two CuO planes within $c_0$.}\BibitemShut
	{Stop}%
	\bibitem [{\citenamefont {Minami}\ \emph {et~al.}(2003)\citenamefont {Minami},
		\citenamefont {Wittorff}, \citenamefont {Yelland}, \citenamefont {Cooper},
		\citenamefont {Changkang},\ and\ \citenamefont {Hodby}}]{minami03}%
	\BibitemOpen
	\bibfield  {author} {\bibinfo {author} {\bibfnamefont {H.}~\bibnamefont
			{Minami}}, \bibinfo {author} {\bibfnamefont {V.~W.}\ \bibnamefont
			{Wittorff}}, \bibinfo {author} {\bibfnamefont {E.~A.}\ \bibnamefont
			{Yelland}}, \bibinfo {author} {\bibfnamefont {J.~R.}\ \bibnamefont {Cooper}},
		\bibinfo {author} {\bibfnamefont {C.}~\bibnamefont {Changkang}}, \ and\
		\bibinfo {author} {\bibfnamefont {J.~W.}\ \bibnamefont {Hodby}},\ }\href
	{\doibase 10.1103/PhysRevB.68.220503} {\bibfield  {journal} {\bibinfo
			{journal} {Phys. Rev. B}\ }\textbf {\bibinfo {volume} {68}},\ \bibinfo
		{pages} {220503} (\bibinfo {year} {2003})}\BibitemShut {NoStop}%
	\bibitem [{\citenamefont {Waske}\ \emph {et~al.}(2007)\citenamefont {Waske},
		\citenamefont {Hess}, \citenamefont {Büchner}, \citenamefont {Hinkov},\ and\
		\citenamefont {Lin}}]{waske07}%
	\BibitemOpen
	\bibfield  {author} {\bibinfo {author} {\bibfnamefont {A.}~\bibnamefont
			{Waske}}, \bibinfo {author} {\bibfnamefont {C.}~\bibnamefont {Hess}},
		\bibinfo {author} {\bibfnamefont {B.}~\bibnamefont {Büchner}}, \bibinfo
		{author} {\bibfnamefont {V.}~\bibnamefont {Hinkov}}, \ and\ \bibinfo {author}
		{\bibfnamefont {C.}~\bibnamefont {Lin}},\ }\href {\doibase
		https://doi.org/10.1016/j.physc.2007.03.169} {\bibfield  {journal} {\bibinfo
			{journal} {Phys. C: Supercond.}\ }\textbf {\bibinfo {volume} {460-462}},\
		\bibinfo {pages} {746} (\bibinfo {year} {2007})}\BibitemShut {NoStop}%
	\bibitem [{\citenamefont {Hartnoll}(2015)}]{hartnoll15}%
	\BibitemOpen
	\bibfield  {author} {\bibinfo {author} {\bibfnamefont {S.~A.}\ \bibnamefont
			{Hartnoll}},\ }\href {\doibase 10.1038/nphys3174} {\bibfield  {journal}
		{\bibinfo  {journal} {Nat. Phys.}\ }\textbf {\bibinfo {volume} {11}},\
		\bibinfo {pages} {54} (\bibinfo {year} {2015})}\BibitemShut {NoStop}%
	\bibitem [{notedata()}]{notedata}%
	\BibitemOpen
	\bibinfo {note} {The data and scripts needed to reproduce the figures can be
		found at: \url{https://doi.org/10.5281/zenodo.6985239}.}\BibitemShut {Stop}%
	\bibitem [{\citenamefont {Shastry}(2008)}]{shastry09}%
	\BibitemOpen
	\bibfield  {author} {\bibinfo {author} {\bibfnamefont {B.~S.}\ \bibnamefont
			{Shastry}},\ }\href {\doibase 10.1088/0034-4885/72/1/016501} {\bibfield
		{journal} {\bibinfo  {journal} {Rep. Prog. Phys.}\ }\textbf {\bibinfo
			{volume} {72}},\ \bibinfo {pages} {016501} (\bibinfo {year}
		{2008})}\BibitemShut {NoStop}%
	\bibitem [{\citenamefont {Dagotto}(1999)}]{dagotto99}%
	\BibitemOpen
	\bibfield  {author} {\bibinfo {author} {\bibfnamefont {E.}~\bibnamefont
			{Dagotto}},\ }\href {\doibase 10.1088/0034-4885/62/11/202} {\bibfield
		{journal} {\bibinfo  {journal} {Rep. Prog. Phys.}\ }\textbf {\bibinfo
			{volume} {62}},\ \bibinfo {pages} {1525} (\bibinfo {year}
		{1999})}\BibitemShut {NoStop}%
	\bibitem [{\citenamefont {Yu}\ \emph {et~al.}(1994)\citenamefont {Yu},
		\citenamefont {Salamon}, \citenamefont {Kopylov}, \citenamefont {Kolesnikov},
		\citenamefont {Duan},\ and\ \citenamefont {Hermann}}]{yu94}%
	\BibitemOpen
	\bibfield  {author} {\bibinfo {author} {\bibfnamefont {F.}~\bibnamefont
			{Yu}}, \bibinfo {author} {\bibfnamefont {M.}~\bibnamefont {Salamon}},
		\bibinfo {author} {\bibfnamefont {V.}~\bibnamefont {Kopylov}}, \bibinfo
		{author} {\bibfnamefont {N.}~\bibnamefont {Kolesnikov}}, \bibinfo {author}
		{\bibfnamefont {H.}~\bibnamefont {Duan}}, \ and\ \bibinfo {author}
		{\bibfnamefont {A.}~\bibnamefont {Hermann}},\ }\href {\doibase
		https://doi.org/10.1016/0921-4534(94)91969-0} {\bibfield  {journal} {\bibinfo
			{journal} {Phys. C: Supercond.}\ }\textbf {\bibinfo {volume} {235-240}},\
		\bibinfo {pages} {1489} (\bibinfo {year} {1994})}\BibitemShut {NoStop}%
	\bibitem [{\citenamefont {Abdel-Jawad}\ \emph {et~al.}(2006)\citenamefont
		{Abdel-Jawad}, \citenamefont {Kennett}, \citenamefont {Balicas},
		\citenamefont {Carrington}, \citenamefont {Mackenzie}, \citenamefont
		{McKenzie},\ and\ \citenamefont {Hussey}}]{abdel06}%
	\BibitemOpen
	\bibfield  {author} {\bibinfo {author} {\bibfnamefont {M.}~\bibnamefont
			{Abdel-Jawad}}, \bibinfo {author} {\bibfnamefont {M.~P.}\ \bibnamefont
			{Kennett}}, \bibinfo {author} {\bibfnamefont {L.}~\bibnamefont {Balicas}},
		\bibinfo {author} {\bibfnamefont {A.}~\bibnamefont {Carrington}}, \bibinfo
		{author} {\bibfnamefont {A.~P.}\ \bibnamefont {Mackenzie}}, \bibinfo {author}
		{\bibfnamefont {R.~H.}\ \bibnamefont {McKenzie}}, \ and\ \bibinfo {author}
		{\bibfnamefont {N.~E.}\ \bibnamefont {Hussey}},\ }\href {\doibase
		10.1038/nphys449} {\bibfield  {journal} {\bibinfo  {journal} {Nat. Phys.}\
		}\textbf {\bibinfo {volume} {2}},\ \bibinfo {pages} {821} (\bibinfo {year}
		{2006})}\BibitemShut {NoStop}%
	\bibitem [{\citenamefont {Abdel-Jawad}\ \emph {et~al.}(2007)\citenamefont
		{Abdel-Jawad}, \citenamefont {Analytis}, \citenamefont {Balicas},
		\citenamefont {Carrington}, \citenamefont {Charmant}, \citenamefont
		{French},\ and\ \citenamefont {Hussey}}]{abdel07}%
	\BibitemOpen
	\bibfield  {author} {\bibinfo {author} {\bibfnamefont {M.}~\bibnamefont
			{Abdel-Jawad}}, \bibinfo {author} {\bibfnamefont {J.~G.}\ \bibnamefont
			{Analytis}}, \bibinfo {author} {\bibfnamefont {L.}~\bibnamefont {Balicas}},
		\bibinfo {author} {\bibfnamefont {A.}~\bibnamefont {Carrington}}, \bibinfo
		{author} {\bibfnamefont {J.~P.~H.}\ \bibnamefont {Charmant}}, \bibinfo
		{author} {\bibfnamefont {M.~M.~J.}\ \bibnamefont {French}}, \ and\ \bibinfo
		{author} {\bibfnamefont {N.~E.}\ \bibnamefont {Hussey}},\ }\href {\doibase
		10.1103/PhysRevLett.99.107002} {\bibfield  {journal} {\bibinfo  {journal}
			{Phys. Rev. Lett.}\ }\textbf {\bibinfo {volume} {99}},\ \bibinfo {pages}
		{107002} (\bibinfo {year} {2007})}\BibitemShut {NoStop}%
	\bibitem [{\citenamefont {French}\ \emph {et~al.}(2009)\citenamefont {French},
		\citenamefont {Analytis}, \citenamefont {Carrington}, \citenamefont
		{Balicas},\ and\ \citenamefont {Hussey}}]{french09}%
	\BibitemOpen
	\bibfield  {author} {\bibinfo {author} {\bibfnamefont {M.~M.~J.}\
			\bibnamefont {French}}, \bibinfo {author} {\bibfnamefont {J.~G.}\
			\bibnamefont {Analytis}}, \bibinfo {author} {\bibfnamefont {A.}~\bibnamefont
			{Carrington}}, \bibinfo {author} {\bibfnamefont {L.}~\bibnamefont {Balicas}},
		\ and\ \bibinfo {author} {\bibfnamefont {N.~E.}\ \bibnamefont {Hussey}},\
	}\href {\doibase 10.1088/1367-2630/11/5/055057} {\bibfield  {journal}
		{\bibinfo  {journal} {New J. Phys.}\ }\textbf {\bibinfo {volume} {11}},\
		\bibinfo {pages} {055057} (\bibinfo {year} {2009})}\BibitemShut {NoStop}%
	\bibitem [{\citenamefont {Yu}\ \emph {et~al.}(1996)\citenamefont {Yu},
		\citenamefont {Kopylov}, \citenamefont {Salamon}, \citenamefont {Kolesnikov},
		\citenamefont {{H}ubbard}, \citenamefont {Duan},\ and\ \citenamefont
		{Hermann}}]{yu96}%
	\BibitemOpen
	\bibfield  {author} {\bibinfo {author} {\bibfnamefont {F.}~\bibnamefont
			{Yu}}, \bibinfo {author} {\bibfnamefont {V.}~\bibnamefont {Kopylov}},
		\bibinfo {author} {\bibfnamefont {M.}~\bibnamefont {Salamon}}, \bibinfo
		{author} {\bibfnamefont {N.}~\bibnamefont {Kolesnikov}}, \bibinfo {author}
		{\bibfnamefont {M.}~\bibnamefont {{H}ubbard}}, \bibinfo {author}
		{\bibfnamefont {H.}~\bibnamefont {Duan}}, \ and\ \bibinfo {author}
		{\bibfnamefont {A.}~\bibnamefont {Hermann}},\ }\href {\doibase
		https://doi.org/10.1016/0921-4534(96)00394-2} {\bibfield  {journal} {\bibinfo
			{journal} {Phys. C: Supercond.}\ }\textbf {\bibinfo {volume} {267}},\
		\bibinfo {pages} {308} (\bibinfo {year} {1996})}\BibitemShut {NoStop}%
	\bibitem [{\citenamefont {Tulipman}\ and\ \citenamefont
		{Berg}(2022)}]{tulipman22x}%
	\BibitemOpen
	\bibfield  {author} {\bibinfo {author} {\bibfnamefont {E.}~\bibnamefont
			{Tulipman}}\ and\ \bibinfo {author} {\bibfnamefont {E.}~\bibnamefont
			{Berg}},\ }\href {https://arxiv.org/abs/2211.00665} {\bibfield  {journal}
		{\bibinfo  {journal} {arXiv:2211.00665 [cond-mat]}\ } (\bibinfo {year}
		{2022})}\BibitemShut {NoStop}%
	\bibitem [{\citenamefont {Bondarenko}\ \emph {et~al.}(2017)\citenamefont
		{Bondarenko}, \citenamefont {Koverya}, \citenamefont {Krevsun},\ and\
		\citenamefont {Link}}]{bondarenko17}%
	\BibitemOpen
	\bibfield  {author} {\bibinfo {author} {\bibfnamefont {S.~I.}\ \bibnamefont
			{Bondarenko}}, \bibinfo {author} {\bibfnamefont {V.~P.}\ \bibnamefont
			{Koverya}}, \bibinfo {author} {\bibfnamefont {A.~V.}\ \bibnamefont
			{Krevsun}}, \ and\ \bibinfo {author} {\bibfnamefont {S.~I.}\ \bibnamefont
			{Link}},\ }\href {\doibase 10.1063/1.5008405} {\bibfield  {journal} {\bibinfo
			{journal} {Low Temp. Phys.}\ }\textbf {\bibinfo {volume} {43}},\ \bibinfo
		{pages} {1125} (\bibinfo {year} {2017})}\BibitemShut {NoStop}%
	\bibitem [{\citenamefont {Varshney}\ \emph {et~al.}(2011)\citenamefont
		{Varshney}, \citenamefont {Yogi}, \citenamefont {Dodiya},\ and\ \citenamefont
		{Mansuri}}]{varshney11}%
	\BibitemOpen
	\bibfield  {author} {\bibinfo {author} {\bibfnamefont {D.}~\bibnamefont
			{Varshney}}, \bibinfo {author} {\bibfnamefont {A.}~\bibnamefont {Yogi}},
		\bibinfo {author} {\bibfnamefont {N.}~\bibnamefont {Dodiya}}, \ and\ \bibinfo
		{author} {\bibfnamefont {I.}~\bibnamefont {Mansuri}},\ }\href {\doibase
		10.4236/jmp.2011.28109} {\bibfield  {journal} {\bibinfo  {journal} {J. Mod.
				Phys.}\ }\textbf {\bibinfo {volume} {2}},\ \bibinfo {pages} {922} (\bibinfo
		{year} {2011})}\BibitemShut {NoStop}%
\end{thebibliography}

%

\end{document}